\documentclass[aps,prb,twocolumn,superscriptaddress,floatfix,longbibliography]{revtex4-2}

\usepackage[utf8]{inputenc}
\usepackage{graphicx}
\usepackage{braket}
\usepackage{bm}
\usepackage{upgreek}
\usepackage{mathtools}
\usepackage{array}
\usepackage{physics}
\usepackage{multirow}

\usepackage{hyperref}

\begin{document}
\title{Energy spectrum of trions in semiconductors}

\author{Pavel A. Belov}
\affiliation{Institut für Physik, Universität Rostock,
  Albert-Einstein-Straße 23-24, 18059 Rostock, Germany}
\email[Email: ]{pavelbelov@gmail.com}
\author{Lucas Happ}
\affiliation{Institut für Quantenphysik and Center for Integrated Quantum Science and Technology (IQST), Universität Ulm, Albert-Einstein-Allee 11, 89081 Ulm, Germany}
\affiliation{Few-body Systems in Physics Laboratory, RIKEN Nishina Center, 2-1 Hirosawa, Wak\={o}, Saitama 351-0198, Japan}
\author{Stefan Scheel}
\affiliation{Institut für Physik, Universität Rostock,
  Albert-Einstein-Straße 23-24, 18059 Rostock, Germany}
\author{Evgeny A. Yarevsky}
\affiliation{Department of Computational Physics, Saint-Petersburg State University,
  Ulyanovskaya 1, 198504 Russia}

\date{\today}

\begin{abstract}
Energy levels of trions in a bulk semiconductor crystal are calculated by two different numerical methods. On the one hand, we construct the finite-element solution of the Schr\"{o}dinger equation in the total angular momentum representation. On the other hand, we use the Gaussian expansion method (GEM), applied to each of the three Faddeev components of the wave function. The results for the total angular momenta $J=0$ and $J=1$ of the trion were obtained for a wide range of the hole-to-electron mass ratios covering most existing semiconductor compounds. The dependencies of the energies on the mass ratio for trions in a bulk crystal qualitatively agree with the results for a simplified model of a trion in a 1D quantum wire, for which the three-body energies can be calculated with arbitrary precision as well as easily approximated using Born-Oppenheimer potential-energy curves. A comparison of our results with data from other publicly available three-body codes as well as with the energies of well-known atomic counterparts, in particular, with vibrational energy levels of the molecular hydrogen ion, shows a good agreement and confirms the accuracy of the numerical data. We found that GEM significantly improves the precision of the obtained trion energies compared to previous calculations. Our calculations demonstrate that, for most semiconductor compounds, only one trion state with zero total angular momentum exists.
\end{abstract}

\maketitle

\section{Introduction}
\label{sec:intro}

Electron-hole bound states such as excitons and more complex three- and four-body states such as 
trions and biexcitons are a key focus of semiconductor 
spectroscopy~\cite{ivchenko,klingshirn2007semiconductor,Kaz14,Plechinger2016,Lin2022,Qi2025,Nguyen2025}.
These few-body systems in confined geometries — particularly in heterostructures and 
transition-metal dichalcogenide (TMDC) monolayers~\cite{Glazov2018,Schmelcher2020} — have 
attracted significant attention due to their potential use in room-temperature excitonic 
devices~\cite{Grosso2009,Andreakou2014,Butov2017} and sensors~\cite{Meyer2020,Chukeev2024}.
This is equally promising for the bulk structures that exhibit excitons with 
large binding energies~\cite{Godde2016,Schweiner16b,Heck17a,Heck25,Jasinski2025,Massignan2025}, 
that are higher than the characteristic energy of lattice thermal vibrations.

Although in conventional bulk semiconductors such as GaAs and CdTe, excitons have relatively 
small binding energies, their ground-state and excited-state peaks can be rather well observed 
experimentally~\cite{ivchenko,Lee1994,Horodysky2006}.
Even smaller binding energies of trions and biexcitons make them more difficult to measure, as 
their diffuse peaks overlap with much sharper excitonic lines.
This is why, despite their long-standing theoretical prediction~\cite{Lampert1958}, the first 
observation of trions (also known as charged excitons) was only possible after the development of 
growth techniques for heterostructures containing quantum wells (QWls)~\cite{Kheng1993}.
Reducing the dimensionality increases the binding energy of electron-hole 
complexes~\cite{Davies1997,Belov2019,Kezerashvili2019}, thus enabling the corresponding peaks 
in the spectra to be resolved experimentally~\cite{Shields1995,Khramtsov2016}.
The large binding energies of electron-hole complexes in bulk Cu$_2$O and ZnO crystals make these 
structures promising candidates for observing trions as well. However, achieving 
uncompensated electrons and holes in these structures remains challenging~\cite{Stolz2026}.

Trions in bulk semiconductors are the simplest solid-state three-body systems.
For bulk materials, many-body effects, giving rise to trions, can be successfully explained in 
terms of quasiparticles (or dressed particles)~\cite{Knox1963}.
Therefore, a convenient framework of effective few-body Hamiltonians, allowing for an 
application of variational methods, is widely used~\cite{Kezerashvili2019}.
However, for example, in TMDC monolayers trions appear at much higher densities of charge 
carriers. For such systems, the properties of trions can only be described from a proper 
many-body point of view, in particular within the Fermi polaron 
approach~\cite{Efimkin2017,Massignan2025,Glazov2026}.
In this context, it is somewhat surprising that the spectra of trions, at least for the simplest 
bulk model of two bands with parabolic dispersions, have not been calculated so far.
The main reason is probably that the three-body problem with the Coulomb interaction allows only 
a numerical solution~\cite{Frolov2001}.
An analytical solution for the ground state of a three-body system is only known for contact 
(zero-range) interactions in one dimension~\cite{McGuire1964}. Short-range and long-range 
interactions between the particles, as well as additional spatial degrees of freedom make the 
problem analytically intractable. In QWls, the situation is even more complicated: one-dimensional confinement breaks the translational invariance and introduces quantization of the 
center-of-mass motion.
Even studying excitons in QWls requires a computationally complex numerical solution to the 
three-dimensional Schr\"{o}dinger equation~\cite{Khramtsov2016,Belov2017,Belov2019,Belov2024}.
Furthermore, due to strong spin-orbit coupling, an accurate theoretical description of exciton 
spectra in bulk crystals requires one to go beyond the simple parabolic band approximation, i.e. 
to take into account the complex valence band 
structure~\cite{Thewes15,Schoene16,Schweiner16a,Schweiner17a,Rommel18,Heck18,Heck25}.
A simplification can be found for very narrow QWls or quantum wires, for which the center-of-mass 
quantization is so strong that the relative motion of the charge carriers can be completely 
separated.
Therefore, even within the simplest two-band model, the trion in a QWl is a multidimensional 
problem that can only be solved by calculating the ground state characteristics using the 
variational approach with predefined wave 
functions~\cite{Stebe1997,vargabook,Usukura1999,Wojs2007}.

In the present article, we study the energy spectra of $X^{+}$ trions, consisting of an electron 
and two holes. The simplest two-band model with parabolic dispersions of the conduction and 
valence bands is assumed. We obtain spectra of trions in both the bulk crystal and the 
one-dimensional geometry (cylindrical quantum wire) and map out their unified dependencies on the 
effective mass ratio. Our study also covers the mass range of the $X^{-}$ trion model, 
which is mathematically identical to the $X^{+}$ model by a formal interchange of electrons and 
holes.
Both $X^{+}$ and $X^{-}$ trion ground states in bulk materials were extensively studied 
theoretically~\cite{Usukura1999,Combescot2019,Marsusi2022}, whereas the existence of excited states is still debated.

The one-dimensional system allows us to accurately determine the structure of energy levels of 
bound states, understand their origins and, therefore, assess the implications for the general 
case without spatial confinement.
Similar one-dimensional three-body models have already been studied before, see 
Refs.~\cite{Mehta2007,Mehta2014,Happ2019,Happ2022}. These works, however, disregarded the 
interaction between heavy particles, considered only zero-range interactions with the light 
particle, and mainly presented ground-state energies. In our article, we consider the long-range 
interactions in all three two-body subsystems and calculate not only the ground state but many 
energy levels of excited states. Thus, in contrast to Refs.~\cite{Semina2008,SemSur2022}, we 
accurately reconstruct the energy spectrum of the one-dimensional three-body system and observe 
how it depends on the mass ratio. The exact energy levels are compared with the 
Born-Oppenheimer~\cite{BO1927} approximation as well.

The rough estimates show that the ground-state binding energy of the three-body system  
with two identical particles, in particular, a trion, weakly depends on the mass ratio of two 
nonidentical constituents: it can be approximated by a constant value for a wide range of mass 
ratios~\cite{Lin1995}.
However, a more precise systematic study by Combescot~\cite{Combescot2019} clearly 
demonstrates nonlinear deviations from a single value.
It considered the trion in a bulk crystal within a two-band model with parabolic 
dispersions of the conduction and valence bands.
The ground state energy has been calculated as a point of the kernel singularity
of the momentum-space integral equation for the scattering amplitude of an electron-exciton 
collision.
A dependence of the ground-state energy on the mass ratio has been obtained. Attention was paid 
to the mass ratios corresponding to the negatively charged hydrogen ion H$^{-}$~\cite{Hill1977} 
and the positively charged hydrogen molecular ion H$_{2}^{+}$~\cite{Landau,Carrington1989}, which 
is a well-studied atomic system.
In particular, the lowest electronic terms of H$_{2}^{+}$, $1s\sigma$ and $2p\sigma$, in 
different approximations including the infinite nuclear mass and neglecting the nuclear repulsion 
were calculated by many authors~\cite{Teller1930,Matsen1953,Bates1953,Peek1965,Slavyanov1976}.
Analytical expressions~\cite{Xie2005} for the electronic potential curves, including the 
well-known Morse potential~\cite{Morse1,Morse2}, for different diatomic molecular ions were 
proposed to facilitate calculation of the vibrational levels~\cite{Wind1965}. Moreover, 
several high-precision variational calculations of the H$_{2}^{+}$ energy levels have been 
reported~\cite{Hilico2000,Korobov2000,Li2007}. The ion H$^{-}$, having only one bound state due 
to a strong Coulomb repulsion of the constituent electrons, has also drawn considerable 
attention~\cite{Hill1977,Petrimoulx2025}. However, an evolution of the three-body spectrum during 
a crossover between the limiting cases of negatively charged hydrogen ion and the positively 
charged hydrogen molecular ion has not been studied so far.

In the present article, we extend the results of Ref.~\cite{Combescot2019} by presenting the full 
spectrum of the bound-state energies and analyzing their dependence on the 
hole-to-electron mass ratio $m_h/m_e$. To this end, the three-body Schr\"{o}dinger 
equation for the total angular momentum values $J=0$ and $J=1$ is numerically solved
by two different variational methods.
The first one is the finite-element solution of the Schr\"{o}dinger equation in the total 
angular momentum representation~\cite{Scrinzi1995,Yarevsky1997EPL,Yarevsky1997,Gagin2009}.
The second method is the Gaussian expansion method, based on the expansion of three 
Faddeev components of the wave function over a basis of Gaussian functions and bipolar 
harmonics~\cite{Hiyama2003,Hiyama2022,happ2026a}.
The obtained dependencies of energy levels on the mass ratio are described and the vibrational 
levels are classified. A formation of three-body excited states, i.e. several lowest vibrational 
modes, for large mass ratios is shown. For the mass ratio corresponding to H$_{2}^{+}$, 20 
calculated vibrational levels for $J=0$ are compared with the high-precision data from 
Refs.~\cite{Hilico2000,Korobov2000}. We also provide data that allow one to approximate the six 
lowest vibrational energies for all reasonable values of the mass ratios, including the limiting 
case of an infinite hole mass. These can be used to tune the parameters of the hole-hole 
('interatomic') potential to significantly facilitate simulation of the three-body system.

In contrast to Refs.~\cite{Filikhin2018,Filikhin2018PLA}, where only the $s$-wave model in 
all subsystems was considered, we take into account the higher partial waves and thus obtain more 
precise results. Our calculations improve precision of the data from 
Refs.~\cite{Usukura1999,Combescot2019,Marsusi2022} and demonstrate that, for most semiconductor 
compounds, only one trion state with zero total angular momentum exists. A single trion state 
exists if the mass ratio of the two different particles, i.e. $m_h/m_e$ in case of $X^{+}$ trion 
and $m_{e}/m_{h}$ for $X^{-}$, is below $9.7$.
For $J=1$, both methods show no trion bound states for mass ratios below $4.2$.

It is worth noting that our numerical solvers are not unique. There are also 
variational~\cite{Stebe1997,vargabook,Usukura1999,Korobov2000,Korobov2018} and 
finite-difference~\cite{Kolganova1998,EDVARDSSON2016} three-body calculations, approaches using 
hyperspherical 
harmonics~\cite{Lin1995,Timofeyuk2012,Timofeyuk2017,Filikhin2018,Kezerashvili2019,Kezerashvili2024}, 
Laguerre polynomial~\cite{Shirokov2009,Caprio2012,Liverts2013} and numerical 
spline-basis~\cite{Bachau2001,Roudnev2005,Belov2013} expansion methods.
However, almost all mentioned works considered mainly the minimum energy quantum state 
and paid little attention to the excited states, i.e., they did not study the entire spectrum 
of the three-body systems.
Nonetheless, some of these algorithms have been released in publicly available three-body 
codes~\cite{Roudnev2012,Liverts2013,Korobov2026}.
In particular, we employed the open-source SWEXPH~\cite{Korobov2026} and 
ThreeBSCbSSR~\cite{Liverts2013} programs in our study for an additional comparison of the 
precision of the ground-state calculations.

In our work, we calculate energies of the bound states by different methods and compare 
the respective results. As our computations were performed for a broad range of effective mass 
ratios, they cover many well-known
semiconductor compounds, for example GaAs, Cu$_{2}$O, and CdTe. Therefore, the obtained results 
can be interesting not only from a purely theoretical point of view, but also for practical 
purposes.
From our calculations one can obtain more precise values of trion binding energies
by scaling the numerical results for particular semiconductor compounds by their corresponding 
exciton Hartree energies.

\section{The model problem of a trion in a quantum wire}

In this section, we consider a model problem of a trion in a cylindrical quantum wire,
which confines the motion of the charge carriers to only one dimension.
This simple model allows for machine-precision solutions for various types of interparticle 
interactions. Its ground state has already been extensively studied in 
Refs.~\cite{Mehta2007,Kartavtsev2009,Mehta2014,Happ2019,Happ2022,SemSur2022}.
Here, we calculate the energy spectrum of bound states and observe its dependence on the masses 
of the constituent particles.
Our results underline the key implications, which are crucial for understanding the behavior of 
the bound-state energies for a general three-body problem, namely for a trion in a bulk crystal. 
For a more detailed justification of the model, including its applicability and relevance to
fabricated semiconductor quantum wires, we refer the reader to Refs.~\cite{Semina2008,SemSur2022}.

\subsection{Formulation of the model problem}
\label{Section2A}

Let us consider the $X^{+}$ trion in a one-dimensional system. For the sake of simplicity, we 
assume parabolic dispersions of the valence and conduction bands with effective masses 
$m_{1}=m_{2}=m_{h}$, $m_{3}=m_{e}$. We also assume strong confinement in the directions 
perpendicular to the quantum wire. Therefore, the motion described by these coordinates can be 
separated. This leads to a simple problem of a one-dimensional motion of the electron and two 
holes over the quantum wire direction, denoted in the following by $z$.
This also makes connection to atomic physics, where a similar model of a one-dimensional 
three-body system was extensively 
studied~\cite{Mehta2007,Kartavtsev2009,Mehta2014,Happ2019,Happ2022}.
We introduce the following dimensionless coordinates: distance is measured in units of the 
two-body (exciton) Bohr radius $a_{B}$, and energy is measured in units of the exciton Rydberg 
energy. The latter is $\text{Ry}_{m_{h}}=\mu e^4 / 2 \varepsilon^2 \hbar^2$, where 
$\mu=m_{e}m_{h}/(m_{e}+m_{h})$ is the reduced mass of the electron-hole subsystem, $e$ is the 
elementary charge and $\hbar$ is Planck's constant. The dielectric constant $\varepsilon$ is a 
common scale parameter which is taken to be 1.
After eliminating the center-of-mass motion of the three-body system, we arrive at the Schr\"{o}dinger equation for the relative motion of the electron and the holes in the quantum wire. The model Hamiltonian reads~\cite{Happ2019,Happ2022}
\begin{align}
\label{eq1}
&  -\frac{2+\sigma}{2 (1+\sigma)}\frac{\partial^{2}}{\partial y^{2}}
-\frac{2\sigma}{(1+\sigma)}\frac{\partial^{2}}{\partial x^{2}} \nonumber\\
& + \frac{2\mu}{\hbar^2}\left[ V_{1}(y-x/2)+V_{2}(y+x/2) +V_{3}(x) \right],
\end{align}
where we denote $\sigma=m_{e}/m_{h}$.
The Jacobi coordinates $x$ and $y$ are the distances between the two holes and between the 
center-of-mass of the holes $(1,2)$ and the electron, respectively~\cite{faddeevbook}.
\begin{figure}
\includegraphics[width=0.95\columnwidth]{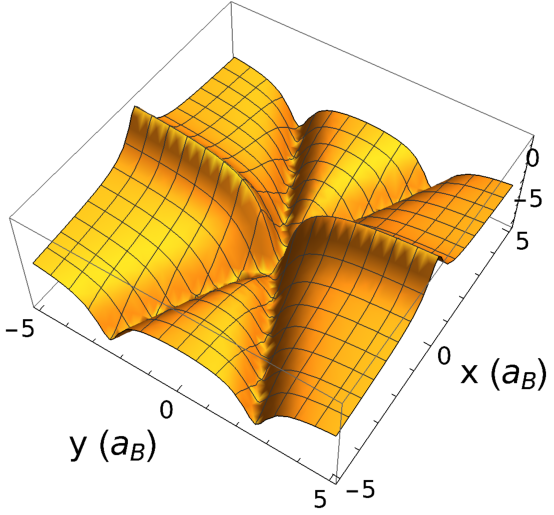}
 \caption{The profile of the three-body potential between particles in the one-dimensional 
 system, i.e. the quantum wire. The energy is measured in units of the exciton Rydberg energy and 
 the coordinates are in units of $a_B$. The parameter of the two-body potential is chosen as 
 $R=0.366$ $a_B$.}
\label{fig:PotentialProfile}
\end{figure}

To simulate the electron-hole interactions in a quantum wire, we use the model attractive 
potential~\cite{Semina2008}
\begin{equation}
\label{eq2}
V_{1,2}^{R}(z) = - \frac{\hbar^2}{2\mu}\frac{2}{\sqrt{z^{2}+R^{2}}}
\end{equation}
which, in contrast to previous studies~\cite{Mehta2014,Happ2019} of contact interactions, is 
long-range. It is obtained by averaging the Coulomb interaction over the radial functions, 
perpendicular to the wire direction~\cite{Elliott1960}.
In general, the parameter $R$ is related to the effective radius of the quantum wire, but it effectively removes the singularity of the Coulomb potential in one dimension.
The hole-hole repulsion is taken into account by a potential of the same form 
$V_{3}^{R}(x)=-V_{1,2}^{R}(x)$. The profile of the resulting total potential in Eq.~\eqref{eq1}
is shown in Fig.~\ref{fig:PotentialProfile}. One can see the valleys of the potential minima 
along the lines $y \pm x/2=0$ as well as two parts of the ridge of the potential maxima along 
the line $x=0$ corresponding to the repulsive interaction of the two holes~\cite{McGuire1964}.

The Schr\"{o}dinger equation with the Hamiltonian~\eqref{eq1} is complemented by appropriate 
boundary conditions. In particular, the trivial (or zero) boundary conditions as 
$x\to \pm \infty$ and $y\to \pm \infty$ are employed for calculating the bound states. As the 
potential~\eqref{eq2} forms a finite QWl for any finite value of $R$, our model equation has, 
at least, one bound state solution~\cite{Chadan2003}.

\subsection{Methods of solution}
\label{sect:MS}

The one-dimensional model~\eqref{eq1} allows for a numerically exact solution, which serves as a 
blueprint for understanding the energy spectrum of the general three-body problem. Moreover, it
can be used for constructing the Born-Oppenheimer (adiabatic) 
potential~\cite{Mustroph2016,Schuurman2018} of electron eigenenergies for fixed positions of the 
holes. This then allows one to solve the one-dimensional equation for the relative motion of the 
holes in the constructed potential, thereby obtaining the three-body bound states within the 
Born-Oppenheimer theory. 

The Born-Oppenheimer approximation is valid when the kinetic energy of the heavy particles, in 
our case the holes, is much smaller than the energy of the heavy-light subsystem~\cite{BO1927}.
This exactly happens in Eq.~\eqref{eq1}, as $m_h \gg m_e$. As $\sigma= m_e / m_h \to 0$, the 
kinetic energy of the relative motion of the heavy particles is $2\sigma$ times smaller than that 
for the heavy-light subsystem. In this case, the energy difference between the Born-Oppenheimer
potential-energy curves $E^{i}_{BO}(x)$, which are obtained from the equation
\begin{align}
\label{eq3}
& \left(-\frac{2+\sigma}{2 (1+\sigma)}\frac{\partial^{2}}{\partial y^{2}} \right. \nonumber\\
& \left. + \frac{2\mu}{\hbar^2}\left[ V_{1}(y-x/2)+V_{2}(y+x/2) +V_{3}(x) \right] \right) 
\psi^{i}(y|x) \nonumber\\
& = E^{i}_{BO}(x) \psi^{i}(y|x)
\end{align}
with $x$ taken as a parameter, is much larger than between the eigenenergies of the 
heavy-particle motion. Then, the Born-Oppenheimer approximation of the three-body energies is 
obtained from a solution of the one-dimensional equation over coordinate $x$ with the particular 
Born-Oppenheimer electron eigenenergy curve $E^{i}_{BO}(x)$ taken as a potential,
\begin{equation}
\nonumber
\left(-\frac{2\sigma}{(1+\sigma)}\frac{\partial^{2}}{\partial x^{2}} + E^{i}_{BO}(x) \right) 
\phi^{i}_{j}(x) = E^{i}_{j} \phi^{i}_{j}(x).
\end{equation}
The opposite limit, $m_h \ll m_e$, i.e. $\sigma= m_e / m_h \to \infty$, can be formally treated 
in a similar way. However, the kinetic energy of the heavy particle (in this case, the electron) 
will be just 4 times smaller. Therefore, the Born-Oppenheimer approach in this case is expected 
to be only approximate.

For arbitrary mass ratio $\sigma$, the Schr\"{o}dinger equation with the Hamiltonian~\eqref{eq1} 
and potentials of the form~\eqref{eq2} can be solved numerically with machine precision. It can 
be done using the standard finite-difference method~\cite{samarskii,Khramtsov2016,Belov2019}, 
more advanced pseudo-spectral method~\cite{Happ2019} or an expansion of the wave function over 
an appropriate basis of polynomials, for example, 
splines~\cite{Bachau2001,Belov2024,Scheuler2024,Aslanidis2025,Kuehner2025}, or finite-element 
functions~\cite{zienkiewicz2006finite,Yarevsky1997}.
In our work, we employed an expansion of the three-body wave function over a basis of cubic 
Hermite splines~\cite{Roudnev2005,Roudnev2011,Belov2013}, which, due to the collocation at the 
Gaussian points~\cite{deBoor1973}, gives a fourth-order accuracy in the step $h$ of equidistant grid, 
i.e. an uncertainty $\sim O(h^4)$. This made it possible to precisely calculate several lowest 
energies of the three-body bound states and to observe their dependence on the mass ratio. The 
size of the calculation domain and the number of collocation points in the numerical algorithm 
were selected in order to achieve the desired level of precision within the considered range of 
mass ratios.

\begin{figure}
\includegraphics[width=0.95\columnwidth]{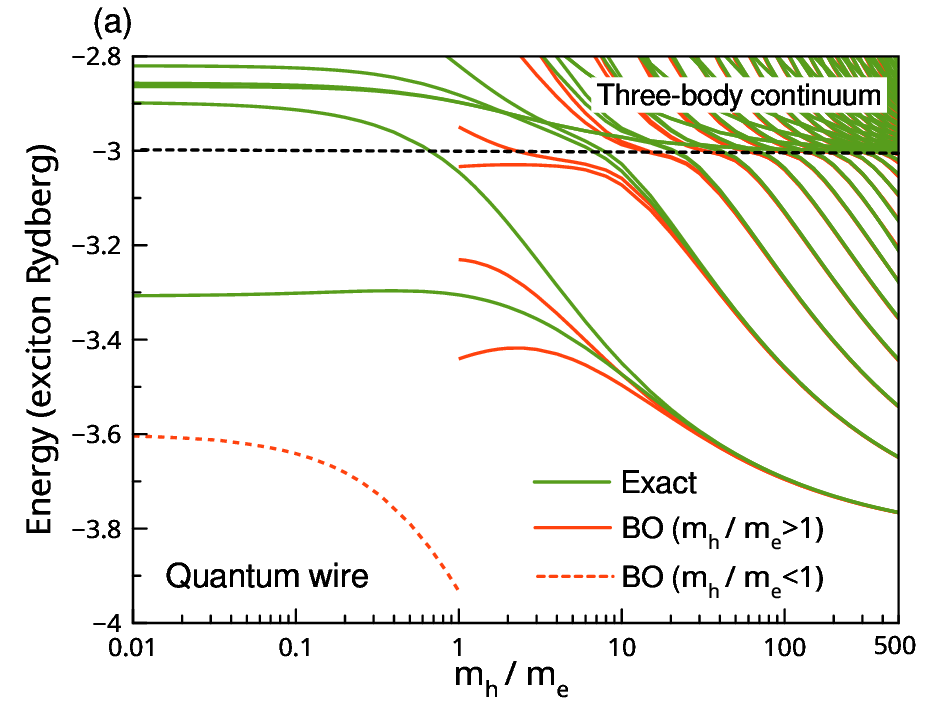}
\includegraphics[width=0.95\columnwidth]{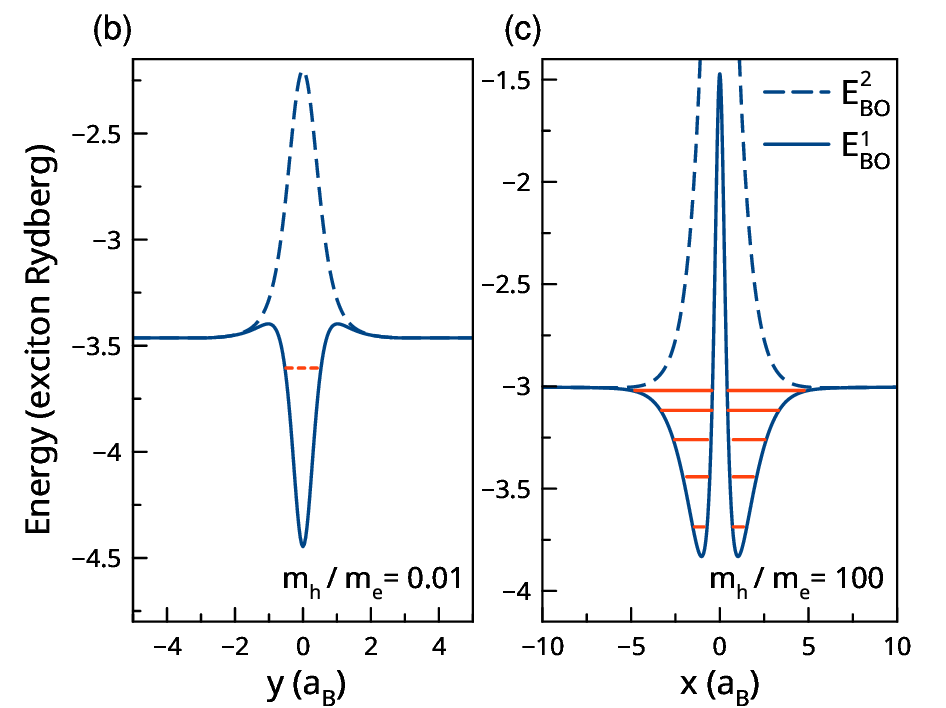}
 \caption{(a) Dependence of the energies of the trion states in a quantum wire on the effective 
 mass ratio $m_h / m_e=\sigma^{-1}$. The green curves show numerically exact results, the red 
 ones are the Born-Oppenheimer (BO) approximations: the dashed curve is a qualitative estimate of the 
 sole energy of the system with two light holes and a heavy electron (similar to H$^{-}$, but 
 with a formal interchange of the electron and holes), whereas the solid ones approach energies 
 of the system with two positively charged heavy constituents (H$_{2}^{+}$-like). The three-body 
 scattering threshold at $-3$~Ry is denoted by the black dashed line. (b) The lowest (solid 
 curve) and upper (dashed curve) potential-energy curves, originating from the Born-Oppenheimer-
 like treatment of the trion in a one-dimensional system, as a function of the distance between 
 center-of-mass of the holes and the electron for the mass ratio $m_{h}/m_{e}=0.01$. The blue 
 curve, i.e. the set of eigenenergies of the light holes for the fixed positions between the 
 electron and the center-of-mass of the holes, represents a single narrow QWl. It is used as an 
 input for Born-Oppenheimer calculation of the sole three-body energy, denoted by dashed red 
 line. (c) The lowest (solid curve) and upper (dashed curve) potential-energy curves, originating 
 from the Born-Oppenheimer-like treatment of the trion in a one-dimensional system, as a function 
 of the distance between two holes for the mass ratio $m_{h}/m_{e}=100$. The blue curve, i.e. the 
 set of eigenenergies of the electron for the fixed distances between the holes, is used as an 
 input for calculation of the three-body energies, i.e. hole-hole vibrational states, denoted by 
 the red color. The potential curve, effectively, represents two weakly-coupled QWls, thus the 
 three-body energy levels are given by pairs, with a splitting of levels within each pair.}
\label{fig:PotentialEnergyCurves}
\end{figure}

\subsection{Results and implications}
\label{Sect2C}

First of all, we separate the calculated bound state energies from resonances. The upper energy 
limit of three-body bound states is determined by the two-body ground state of the 
one-dimensional potential~\eqref{eq2}. This energy of an electron-hole pair (exciton) in a 
quantum wire is the lowest scattering threshold and defines the lower boundary of the three-body 
continuum. As we are free to take any value of the parameter $R$ from a reasonable range, we 
have chosen it to be $R=0.366 \, a_B$ in order to obtain the two-body ground state energy of 
almost exactly $-3$~Ry and thus to simplify the analysis of the results. For a detailed 
justification of the choice of the parameter $R$, see Refs.~\cite{Semina2008,SemSur2022}.

We calculated the energies of the $X^{+}$ trion in a quantum wire and observed their dependence 
on the effective mass ratio $m_h / m_e$. The obtained energies in the units of the exciton
Rydberg are shown in Fig.~\ref{fig:PotentialEnergyCurves}(a). The green curves are the 
numerically exact results, whereas the dashed and solid red curves show the Born-Oppenheimer 
approximations. One can see that in both cases the number of bound states increases with growing 
mass ratio. For $m_{h}/m_{e} \lesssim 0.7$ there is only one bound state. In this sense, the 
system is similar to the negatively charged hydrogen ion, H$^{-}$. The rigorous proof of the 
fact that H$^{-}$ has only one bound state has been given in Ref.~\cite{Hill1977}.

In our case, the Born-Oppenheimer approximation can be used to qualitatively explain the 
emergence of the only bound state for small values of the mass ratio, although its quantitative 
predictive power is still quite moderate for $m_{h}/m_{e} \ll 1$.
The point is that the mass-dependent factors in the kinetic terms in Eq.~\eqref{eq1}, are of the same order: one factor, i.e. for the kinetic term of the light-particle relative motion, is only 4 times larger than the other.
Nevertheless, it means that the motion of the light particles (in this case, holes) generally 
produces larger energies and it can be formally considered as a faster subsystem.
Thus, one can construct the Born-Oppenheimer potential-energy curves of the hole-hole motion by 
taking $y$ as a parameter and averaging over $x$, in a similar way as described in 
Sec.~\ref{sect:MS}.
The analysis shows that the lowest Born-Oppenheimer potential curve obtained in this way forms 
a one-dimensional single QWl. Therefore, it can produce a finite number 
of three-body bound states~\cite{Yang1989,Chadan2003}. However, this QWl is quite narrow and, in 
fact, accommodates only one energy level, see Fig.~\ref{fig:PotentialEnergyCurves}(b).
The QWl width is related to a repulsion of the charge carriers represented, in turn, by the 
magnitude of the potential profile ridge in Fig.~\ref{fig:PotentialProfile}. When averaging over 
$x$, this ridge contributes to the barriers of the Born-Oppenheimer potential. As $R$ decreases, 
this ridge becomes higher and sharper. This causes the QWl to be deeper and narrower.
In fact, the thickness of the QWl is less related to the variation of $R$, rather than the depth. 
Therefore, as $R\to 0$ the QWl becomes much deeper, while its thickness decreases not as 
dramatically. As a result, not only for a given value of $R=0.366$~$a_B$, but for all smaller 
ones, leading asymptotically to the Coulomb potential, the system cannot accommodate more than 
one bound state.
For $m_{h}/m_{e} \lesssim 1$, the Born-Oppenheimer approximation gives only a qualitative 
description, namely the correct number of bound states, but the obtained energy values 
noticeably differ from the numerically exact ones. The coefficients 
in the kinetic terms in Eq.~\eqref{eq1} only slightly vary for $m_{h}/m_{e} \lesssim 0.1$. 
Therefore, for such a range of mass ratios the energies in Fig.~\ref{fig:PotentialEnergyCurves}
(a) almost do not change and look like horizontal lines.

For $m_{h}/m_{e} \gtrsim 0.7$, there are at least two bound states. As $m_{h}/m_{e}$ grows 
further, the number of bound states increases: the energy levels gradually drop down from the 
continuum into the discrete part of the spectrum. The three-body energy levels are given by 
pairs, with a fine splitting of levels within each pair for $m_h \gg m_e$. These energy levels 
are analogues of the vibrational energies of the diatomic molecule~\cite{Landau}.
It is natural that the Born-Oppenheimer approximation should be valid for such mass ratios as 
described in Sect.~\ref{sect:MS}. The green and red curves overlap better as mass ratio grows.
Interestingly, for $m_{h}/m_{e} \gtrsim 1$ the Born-Oppenheimer approximation overestimates the 
binding energies of the two lowest energies by 4~\% and 6~\%. A similar overestimation was reported for zero-range 
potentials in Ref.~\cite{Happ2019}. It is related to the fact that the Born-Oppenheimer approach 
only gives a lower bound to the energy~\cite{Hill1977}.

The splitting of the three-body energy levels into pairs can be understood from the plot of 
Born-Oppenheimer electron potential-energy curves in Fig.~\ref{fig:PotentialEnergyCurves}(c).
The electron potential curves are obtained from Eq.~\eqref{eq3}, i.e. by averaging over $y$.
Due to such averaging of the potential profile, the lowest potential curve $E^{1}_{BO}$ has two 
QWls separated by a narrow barrier, associated with the hole repulsion.
If the separating barrier were infinite, the wave functions in each QWl would be independent and 
thus the energy levels would be degenerate. As the barrier is finite, the wave functions 
originating from each QWl are coupled and the three-body energy levels are split. This becomes 
more pronounced for a mass ratio $m_{h}/m_{e}\sim 1$, for which the ground state wave functions 
for each QWl are most extended. For larger mass ratios, the ground state wave functions become 
more squeezed and their overlap is decreasing. The excited states have different nodal regions of 
opposite signs of the wave function. Therefore, their coupling and level splitting is much 
smaller. Analysis shows that the first excited energy curve $E^{2}_{BO}$ has two very shallow 
minima which can produce very weakly bound states, unable to be resolved with a reasonable 
precision.

\emph{Implications.}
The main result of this model study is that the increase in the mass ratio 
$m_h / m_e=\sigma^{-1}$ leads to an increase in the number of bound states.
The limit $m_h / m_e \to 0$ exhibits only one bound state, as predicted for the analogous atomic 
H$^{-}$ system. With the growth of $m_h / m_e$, more bound states appear. This is related to the 
decrease $\sim\sigma$ of the kinetic energy of relative hole-hole motion as 
$m_h / m_e \to \infty$. As a result, more three-body bound states originating from the hole-hole 
vibrations appear in the spectrum.
Their energies are simultaneously decreasing, gradually filling the wells of the Born-Oppenheimer
potential-energy curves. In the limit of $m_h / m_e  = \sigma^{-1} = \infty$ the second kinetic 
term in Eq.~\eqref{eq1} vanishes, meaning that the quantization disappears and the density of 
three-body states forms a continuum above the minimum of the energy curve. A similar behavior 
should be observed in the general case of a trion in a bulk semiconductor crystal.

\section{Trion in a bulk crystal}

In this section, we consider a three-body Coulomb system in three dimensions. A particular 
example of this in solid-state physics is the trion in a bulk semiconductor crystal.
In atomic physics, the trion's counterparts are the negatively charged hydrogen ion 
H$^{-}$~\cite{Hill1977} and the positively charged hydrogen molecular ion 
H$_2^{+}$~\cite{Landau,Carrington1989}.
We further describe two distinct methods which are used to solve the three-body problem 
and obtain the energies of the trion. Although both methods are variational, they are based
on completely different expansions of the three-body wave function. Thus, the outcomes of these 
methods complement one another and allow a more detailed comparison and verification of the 
results.

\subsection{Formulation of the general three-body problem}

We consider the Hamiltonian of three spinless particles $(m_1,m_2,m_3)$ that interact 
via pairwise potentials. It is assumed that each of the potentials depends only on the radius 
vector between two particles. Due to translational invariance, the center-of-mass motion of 
such a three-body system can be eliminated. The Hamiltonian of the relative motion of 
these particles in the three-dimensional space is conventionally written in Jacobi 
coordinates~\cite{faddeevbook} as
\begin{align}
\label{eq3DH}
   & -\frac{\hbar^2}{2 \mu_{12,3}} \Delta_{\mathbf{r}_{12,3}}
    -\frac{\hbar^2}{2 \mu_{12}} \Delta_{\mathbf{r}_{12}} \nonumber \\
   & +V_{1}\left( \mathbf{r}_{12,3} - \frac{m_1}{m_1+m_2} \mathbf{r}_{12} \right)
    +V_{2}\left( \mathbf{r}_{12,3} + \frac{m_2}{m_1+m_2} \mathbf{r}_{12} \right)  \nonumber \\
   & +V_{3}\left( \mathbf{r}_{12} \right).
\end{align}
Here the reduced masses are given by the masses of the particles: $1/\mu_{12}=1/m_{1}+1/m_{2}$ 
and $1/\mu_{12,3}=1/(m_{1}+m_{2})+1/m_{3}$.
The terms $V_{i}$
define interactions between particles of adjacent pairs $(j,k)$, where indices $\{i,j,k\}$ are cyclic permutations of $\{1,2,3\}$.
The vector $\mathbf{r}_{12}$ specifies the relative motion of the particles in the pair $(1,2)$ 
and the vector $\mathbf{r}_{12,3}$ defines the motion of the particle $3$ with respect to the 
center-of-mass of the pair $(1,2)$.

The Schr\"{o}dinger equation with the Hamiltonian~\eqref{eq3DH} has to be complemented by 
appropriate boundary conditions: vanishing wave functions are used for calculation of the bound state solutions, whereas scattering boundary conditions can be used for constructing the quantum scattering problem~\cite{faddeevbook}.
Note, that an alternative formulation of the scattering problem in terms of the Faddeev components of the wave function significantly simplifies its formulation and numerical solution in configuration space~\cite{Merkuriev1976}.
As we will show below, the decomposition of the wave function into the Faddeev components also serves to efficiently solve the bound state problem.

\subsection{Method of solution: Total angular momentum representation}
\label{sec:methodFEM}

In this section, we describe our first approach to calculate energies of the trion states.
It is based on the total angular momentum representation of the three-body problem and described in detail in Refs.~\cite{Yarevsky1997EPL,Yarevsky1997}.
By expanding the three-body wave function over the Wigner D-functions, it allows a reduction of the initial six-dimensional Hamiltonian~\eqref{eq3DH} to a system of coupled three-dimensional equations for the part of the wave function describing the relative motion of the particles within the three-body plane.
These equations can be further effectively solved using the finite-element method.

\subsubsection{Wigner D-function expansion}

The quantum three-body system is characterized by a total angular momentum $J$ and its projection 
$M$. The essence is that the three-body wave function $\Psi^{J}_{M}(\mathbf{R})$
can be represented in terms of Wigner D-functions~\cite{Varshalovich1988,Abramov2008} as
\begin{align}
\label{eq4}
    & \Psi^{J}_{M}(\mathbf{r}_{12},\mathbf{r}_{12,3})=\sum_{M'=0}^{J} \frac{1}{\sqrt{2+2\delta_{M'0}}}  \nonumber \\
    & \left( D^{J}_{M,M'}(\alpha,\beta,\gamma) +(-1)^{M'} D^{J}_{M,-M'}(\alpha,\beta,\gamma) \right) \psi^{J}_{M'} (x,y,\theta).
\end{align}
Here $(\mathbf{r}_{12},\mathbf{r}_{12,3})$ is a pair of Jacobi vectors.
The angles $(\alpha,\beta,\gamma)$ are the Euler angles defining the three-body plane, and 
$(x,y,\theta)$ are the coordinates of the particles in this plane. As in Sec.~\ref{Section2A}, 
the distance between particles $1$ and $2$ is denoted as $x=|\mathbf{r}_{12}|$, whereas the 
distance between particle $3$ and the center-of-mass pair $(1,2)$ is denoted by 
$y=|\mathbf{r}_{12,3}|$. The angle $\theta$ is between $\mathbf{r}_{12}$ and $\mathbf{r}_{12,3}$.

Using the expansion~\eqref{eq4} in the three-body Schr\"{o}dinger equation with the Hamiltonian~\eqref{eq3DH} leads to the system of equations
\begin{equation}
\label{eq5}
    \sum_{M'=0}^{J} \left( H^{J}_{MM'} - E \delta_{MM'} \right) \psi^{J}_{M'} (x,y,\theta) = 0, \quad M=0,\ldots, J
\end{equation}
for the radial part $\psi^{J}_{M} (x,y,\theta)$ of the wave function, describing the 
relative motion of the particles in the three-body plane. The solution of this 
system gives the components of the wave functions for all projections $M$ of a predefined 
total angular momentum $J$. In Eq.~\eqref{eq5}, the diagonal matrix elements of the 
tridiagonal Hamiltonian matrix are
\begin{align*}
    & H^{J}_{MM} = \nonumber \\
    & - \frac{\hbar^2}{2\mu_{12}} \frac{1}{x} \frac{\partial^2}{\partial x^2} x
    - \frac{\hbar^2}{2\mu_{12,3}} \left[ \frac{1}{y} \frac{\partial^2}{\partial y^2} y - \frac{J(J+1)-2M^2}{y^2} \right] \nonumber \\
    & - \frac{\hbar^2}{2} \left( \frac{1}{\mu_{12}x^2} + \frac{1}{\mu_{12,3}y^2} \right)
    \left( \frac{\partial^2}{\partial \theta^2} + \cot{\theta} \frac{\partial}{\partial \theta} - \frac{M^{2}}{\sin^2{\theta}} \right)  \nonumber \\
    & + V(x,y,\theta).
\end{align*}
By defining $\lambda_{\pm}(J,M)=\sqrt{J(J+1)-M(M \pm 1)}$,
the upper- and lower-diagonal matrix elements read as
\begin{align*}
    & H^{J}_{M \; M\pm1} = \nonumber \\
    & - i \hbar^2 \sqrt{1+\delta_{M,0}\delta_{M\pm 1,1}+\delta_{M,1}\delta_{M\pm 1,0}}\frac{\lambda_{\pm}(J,M)}{2\mu_{12,3}y^2} \nonumber \\
    & \times \left( \frac{\partial}{\partial \theta} + (1\pm M) \cot{\theta} \right).
\end{align*}
Note that $\psi^{J}_{-1} (x,y,\theta) \equiv 0$.

The potential $V(x,y,\theta)$ defines the interactions, i.e. the nature of the three-body system. For the 
$X^{+}$ trion under consideration, it is given by a sum of three pairwise Coulomb potentials:
\begin{equation}
\nonumber
    V(x,y,\theta)= \frac{e^2}{\varepsilon} \left(-\frac{1}{|\mathbf{r}_{23}|}-\frac{1}{|\mathbf{r}_{13}|}+\frac{1}{|\mathbf{r}_{12}|} \right),
\end{equation}
where $e$ is the elementary charge and $\varepsilon$ is the dielectric constant taken to be 1. 
Due to the equal effective masses of the two holes, $m_{1}=m_{2}=m_{h}$, the distances are 
explicitly given as $|\mathbf{r}_{13}|^2=x^2/4+y^2-xy \cos{\theta}$ and 
$|\mathbf{r}_{23}|^2=x^2/4+y^2+xy \cos{\theta}$.

\subsubsection{Numerical solution}
\label{NumSolution}
The numerical solution of Eq.~\eqref{eq5} for a predefined total angular momentum $J$ has been 
performed using the finite-element three-body solver, described in detail in 
Refs.~\cite{Scrinzi1995,Yarevsky1997EPL,Yarevsky1997}. Here we only briefly sketch the basic ideas.

The main advantage of the total angular momentum representation is that the structure of 
Eq.~\eqref{eq5} implies coupling only three components of the wave function: those with $M'=M$ 
and $M'=M\pm 1$. This leads to a block-tridiagonal form of the Hamiltonian matrix, which 
significantly reduces the numerical complexity~\cite{Belov2017b}.
The computational domain over coordinates $(x,y,\cos{\theta})$ consists of a rectangular cuboid, 
divided into a finite number of elements (FEM). The component $\psi^{J}_{M}$ of the total wave 
function is expanded into a sum of local basis functions as
\begin{equation}
\label{eqExpansion}
    \psi^{J}_{M} (x,y,\cos{\theta}) = \sum_{i,m} c^{JM}_{im} f^{JM}_{im} (x,y,\cos{\theta}).
\end{equation}
The terms $c^{JM}_{im}$ are expansion coefficients defined at the particular element $i$, whereas 
the index $m$ specifies the local basis function at this element.
The basis functions over the angular variable are the associated Legendre polynomials. For the 
coordinate variables, the Legendre polynomials are multiplied by damping exponentials properly 
chosen to increase the convergence of the bound state solutions.
Using the expansion~\eqref{eqExpansion} in the Schr\"{o}dinger equation~\eqref{eq5} leads to the 
generalized eigenvalue problem of the form
$\mathcal{H}\mathbf{c}=E\mathcal{S}\mathbf{c}$,
where the vector $\mathbf{c}$ consists of the expansion coefficients.

As already mentioned, for an arbitrary total angular momentum $J$, the Hamiltonian matrix 
$\mathcal{H}$ is block-tridiagonal. Moreover, as each block of the Hamiltonian matrix 
$\mathcal{H}$ and the overlap matrix $\mathcal{S}$ are sparse, an efficient numerical solution 
of the problem can be obtained using the iterative Arnoldi algorithm, realized in 
ARPACK~\cite{arpackuserguide}. The linear system at each iteration of the algorithm is solved 
using the PARDISO library~\cite{Schenk2004}, implemented in Intel MKL~\cite{mkl}.

\subsection{Method of solution: Gaussian Expansion Method}
\label{sec:methodGEM}

In this section we describe our second approach, the Gaussian Expansion Method (GEM), which is 
used as an alternative for calculating the trion bound-state energies. In this approach, the 
three-body wave function is expanded in a large, non-orthogonal basis of Gaussians which allows a 
very efficient computation of the bound states.
This variational method is described in detail in Ref.~\cite{Hiyama2003} and we use its 
implementation in the open-source package FewBodyToolkit.jl~\cite{happ2026a} developed by one of 
the authors.

\begin{figure}[htbp]
    \centering
    \includegraphics[width=1.0\linewidth]{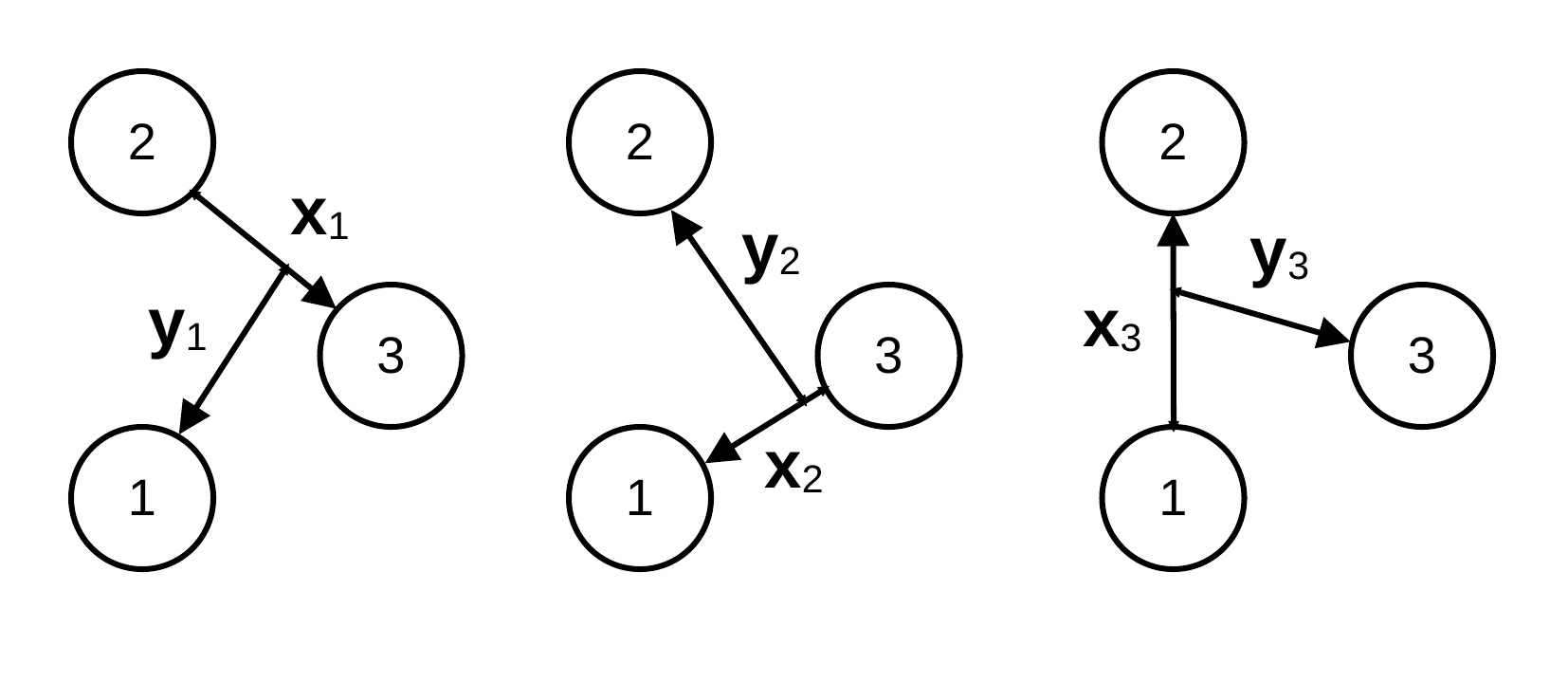}
    \caption{Three sets of Jacobi vectors $(\mathbf{x}_j,\mathbf{y}_j)$, $j=1,2,3$.}
    \label{fig:jacobisets}
\end{figure}

\subsubsection{Jacobi vectors and Faddeev components}

The relative motion of three particles is described using two spatial vectors, in particular 
Jacobi vectors defined by corresponding coordinates. For a three-body system, there are three
alternative ways of choosing these vectors, see Fig.~\ref{fig:jacobisets}.
Let us, for definiteness, consider the first pair of vectors $(\mathbf{x}_1,\mathbf{y}_1)$.
The three-body wave function $\Psi^{J}_{M}(\mathbf{x}_1,\mathbf{y}_1)$ can be represented by a sum~\cite{faddeevbook,Hiyama2003}
\begin{equation}
  \Psi^{J}_{M}(\mathbf{x}_1,\mathbf{y}_1) = \sum_{j=1}^{3}\mathcal{F}^{(j)J}_{M}\left(\mathbf{x}_j,\mathbf{y}_j\right),
  \label{eq:faddeev}
\end{equation}
of three so-called Faddeev components $\mathcal{F}^{(j)J}_{M}$, each of which is expressed in 
terms of a corresponding pair of Jacobi coordinates $(\mathbf{x}_j$, $\mathbf{y}_j)$.
Such a decomposition in all three sets simultaneously allows to describe each pairwise Coulomb 
interaction in its most suited coordinate set. This is important not only for scattering 
problems in order to properly treat the complicated boundary conditions of the three-body 
collisions but, as we will show below, also for calculation of the bound states.
We note that only the three-body bound-state wave function is normalized to unity, not each of 
the Faddeev components individually.

In Eq.~\eqref{eq:faddeev}, the Jacobi vectors representing the second and the third Faddeev 
components are linear combinations of vectors $(\mathbf{x}_1,\mathbf{y}_1)$~\cite{faddeevbook}.
The fact that the sets of Jacobi coordinates are not independent, provides a key advantage of the 
current approach. The transformation between the Jacobi sets affects the representation of the 
Faddeev component (and, more generally, the wave function):
for instance, a component given by pure $s$~waves in one Jacobi set is not given by the $s$~waves 
in the others. Instead, it includes contributions from all partial waves.
The representation~\eqref{eq:faddeev} using all Faddeev components and employing only $s$-wave 
contributions ($l=L=0$) in each pair of the coordinates ($\mathbf{x}_j$, $\mathbf{y}_j$), $j=1,2,3$,
therefore already implicitly contains a large amount of angular correlations in the total wave 
function. Thus, this three-set representation is usually good enough for approximation of the 
wave function even for a relatively small maximal number of partial waves $l_{\max}$ and 
$L_{\max}$ per set. The main difficulty of this representation originates from the transformation 
between the Jacobi coordinate sets. As it will be shown in Sect.~\ref{SectGBF}, using Gaussian 
basis functions allows one to overcome this issue and keep calculations efficient.

For the particular case of the trion, the representation~\eqref{eq:faddeev} is simplified due to 
the identity of two out of three particles.
The symmetry of the Hamiltonian under exchange of the two identical particles leads to an 
equivalence of two Faddeev components. Hence, it is sufficient to describe the system by only 
two components in Eq.~\eqref{eq:faddeev}: one where the pair consists of two identical particles, 
and the other where the pair is built from two different particles.
Effectively, this reduces the Schr\"odinger equation, from a $3\times3$ matrix-type eigenvalue 
equation to a $2\times2$ one.

\subsubsection{Gaussian basis functions}
\label{SectGBF}

To solve the problem, each Faddeev component ($j=1,2,3$) in its proper Jacobi coordinates is expanded into a linear combination~\cite{Hiyama2003}
\begin{align}
  \mathcal{F}_M^{(j)J}(\mathbf{x}_j,\mathbf{y}_j)
  = \sum_{\alpha} C^{(j)}_{\alpha}
    \sum_{m_l,M_L} & \langle l\,m_l\,L\,M_L \,|\, J\,M\rangle\, \nonumber \\
    & \times \phi_{nlm_l}(\mathbf{x}_j)\,\Phi_{NLM_L}(\mathbf{y}_j),
  \label{eq:faddeev_full}
\end{align}
of basis functions, which are chosen to be the products of Gaussians and the solid harmonics~\cite{Varshalovich1988}:
\begin{align}
  \phi_{nlm_l}(\mathbf{x}) &= {\mathcal N}_{nl}\, e^{-\nu_n x^{2}}\, x^{l} Y_{lm_l}(\hat{\mathbf{x}}), \label{eq:gauss1}\\
  \Phi_{NLM_L}(\mathbf{y}) &= {\mathcal N}_{NL}\, e^{-\lambda_N y^{2}}\,  y^{L} Y_{LM_L}(\hat{\mathbf{y}}).
  \label{eq:gauss2}
\end{align}
The multi-index $\alpha$ runs over $\{n,l,N,L\}$ describing the angular momenta $l$, $L$ and the 
Gaussian-range indices $n$, $N$ for the coordinates $\mathbf{x}$ and $\mathbf{y}$, respectively. 
The Clebsch-Gordan-coefficients $\langle l\,m_l\,L\,M_L \,|\, J\,M\rangle$ couple $l$ and $L$ to 
a fixed $J$ and $M$.
This angular expansion is equivalent to bipolar harmonics~\cite{Varshalovich1988,Korobov2026}.
The coefficients ${\mathcal N}_{nl}$, ${\mathcal N}_{NL}$ are normalization factors.

The parameters $\nu_{n}$ and $\lambda_{N}$ define the widths of the Gaussian functions. Choosing 
them in a geometric progression
\begin{align*}
  \nu_n = 1/x_n^2, \qquad
  x_{n} = x_{1} \times \left(\frac{x_{n_{\max}}}{x_1}\right)^{\frac{n-1}{n_{\max}-1}}, \\
  \lambda_N = 1/y_N^2, \qquad
  y_{N} = y_{1}\times \left(\frac{y_{N_{\max}}}{y_1}\right)^{\frac{N-1}{N_{\max}-1}},
  \label{eq:geom2}
\end{align*}
allows an accurate description of each Faddeev component and the entire three-body wave function 
both at the short-range and the long-range asymptotic domain, while reducing the variational 
problem to only three parameters per coordinate ($n_{\max}$, $x_1$, $x_{n_{\max}}$, and 
$N_{\max}$, $y_1$, $y_{N_{\max}}$, respectively).

To facilitate an efficient approximation of the intermediate-range behavior of the 
highly-excited, strongly oscillatory states, complex-ranged Gaussians are employed.
They are obtained by the substitution~\cite{Hiyama2003}
\begin{equation}
  \nu_n \to \nu_n \times (1 \pm i\,\omega),\qquad \lambda_N \to \lambda_N \times (1 \pm i\,\omega),
  \label{eq:cr}
\end{equation}
in Eqs.~\eqref{eq:gauss1} and ~\eqref{eq:gauss2}. To ensure that the total basis can still 
represent a real wave function, basis functions associated with both signs are used. The 
resulting functions are damped oscillations which reproduce the nodal structure of excited states 
more accurately than the original real Gaussians.
Since Eq.~\eqref{eq:cr} keeps the Gaussian functional form unchanged, the same analytical 
expressions for matrix elements can be applied. 

A fast computation of matrix elements for higher partial waves requires an additional 
modification of the basis functions. Since each Faddeev component is expanded in a basis defined 
in its proper coordinate set, evaluating the matrix elements involves transforming the basis 
between the different Jacobi sets.  For $s$~waves this is straightforward, as the product of 
Gaussians in $(\mathbf{x}_j, \mathbf{y}_j)$ remains Gaussian, 
$\sim \exp\left(-\mathbf{v}^T \mathcal{M} \mathbf{v}\right)$, 
with $\mathbf{v} = (\mathbf{x}_k, \mathbf{y}_k)$. The non-diagonal $2\times2$ matrix 
$\mathcal{M}$ contains the information about the widths of the original Gaussians and the masses 
of the particles.

For higher partial waves $l,L > 0$, the solid harmonics, i.e. the vector-argument functions 
$x^l Y_{lm_l}(\hat{\mathbf{x}})$ and $y^L Y_{LM_L}(\hat{\mathbf{y}})$, in Eqs.~\eqref{eq:gauss1} 
and~\eqref{eq:gauss2} make the transformation within their arguments laborious.
To avoid this, the infinitesimally shifted Gaussian lobe (ISGL) functions~\cite{Hiyama2003} are 
used. The main idea is to represent the solid harmonics by finite superpositions of Gaussians 
displaced by infinitesimally shifted vectors. This allows one to perform the transformation 
between the Jacobi coordinate sets in these shifted Gaussians, that is much more straightforward 
than for the solid harmonics. The infinitesimal limit to zero can then be taken analytically and, 
since every integral is then of purely Gaussian type, enables finding expressions for the matrix 
elements in closed form for arbitrary $l$ and $L$.
In general, this depends on the choice of potential. For the Coulomb potential considered here 
and several other potentials, closed-form expressions for matrix elements have been 
obtained~\cite{Hiyama2018}.

The angular momenta for the two relative coordinates are coupled to the total angular momentum 
as $\mathbf{J} = \mathbf{l} + \mathbf{L}$, which restricts possible $(l,L)$ channels via the 
triangle condition $|l-L|\leq J \leq l+L$ and the parity $(-1)^{l+L} = \pm 1$. This is the case 
for each of the three Jacobi sets. Then, for $J=0$, the same angular basis for $\mathbf{x}_j$ 
and $\mathbf{y}_j$, $j=1,2,3$, can be used, $l=L$. For $J=1$, the sum $l+L$ is restricted to be 
odd and hence $|l-L|=1$. For the component whose pair has two identical particles, the exchange 
symmetry constrains $l$ further: to even values for $J=0$, and to odd values for $J=1$.

\subsubsection{Numerical solution}

As the Gaussian basis functions are non-orthogonal, the variational principle leads to a 
generalized eigenvalue problem $\mathcal{H}\mathbf{c}=E\mathcal{S}\mathbf{c}$ with a dense 
overlap matrix $\mathcal{S}$. A geometric range grid, dense enough to converge the energy, can 
contain nearly linearly-dependent functions, which make $\mathcal{S}$ ill-conditioned. Its 
smallest eigenvalues reach the level of the working precision and may even become slightly 
negative due to rounding errors. We therefore first diagonalize $\mathcal{S}$ and discard all 
directions whose eigenvalues fall below a relative threshold of $10^{-14}$ of the largest one by 
projecting $\mathcal{H}$ onto the surviving subspace prior to the final diagonalization.
For the basis sizes used in our calculations, this removes between $4\%$ and $21\%$ of the 
nominal dimensions while keeping the converged energies unchanged.

The numerical parameters used in our calculations are summarized in Table~\ref{tab:numparams_GEM}.
The angular basis is truncated at $l\le l_{\max}$ and $L \le L_{\max}$.
The corresponding matrix sizes are $N_{\mathrm{basis}} = 4\,n_{\max} N_{\max} n_{lL}$. Here the 
factor of $4$ is due to the complex-ranged Gaussians of both signs over each radial coordinate, 
and $n_{lL}$ is the number of $(l,L)$ channels taken into account, summed, for the trion, over 
the two Faddeev components. For example, there are 2 channels, $(0,0)$ and $(2,2)$, for the component 
whose pair has two identical particles and additional 3 channels, $(0,0)$, $(1,1)$, and $(2,2)$, 
for another component in case of $J=0$ and $L_{\max}=l_{\max}=2$. The complex-ranged Gaussians 
were obtained by the parameter $\omega = 0.9$ in Eq.~\eqref{eq:cr}.

The numerical algorithm is implemented in the open-source package 
FewBodyToolkit.jl~\cite{happ2026a}, which takes advantage of the linear algebra functions in the Julia programming language largely calling the original functions from LAPACK~\cite{LAPACK}.
\begin{table}[htbp]
    \centering
    \begin{tabular}{lcccccccc}
    \hline
    \hline
    Configuration & $n_{\max}$ & $N_{\max}$ & $l_{\max}$ & $x_1$ & $x_{n_{\max}}$ & $y_1$ & $y_{N_{\max}}$ & $N_{\mathrm{basis}}$ \\
    \hline \hline
    $J=0$, $r < 0.5$   & 30 & 22 & 2 & 0.003 & 10 & 0.5 & 15 & 13\,200 \\
    $J=0$, $r \ge 0.5$ & 30 & 22 & 2 & 0.005 & 5  & 0.5 & 15 & 13\,200 \\
    $J=0$, H$_2^+$  & 36 & 26 & 2 & 0.002 & 5  & 0.5 & 15 & 18\,720 \\
    $J=0$, H$^-$    & 36 & 26 & 2 & 0.002 & 10 & 0.5 & 15 & 18\,720 \\
    $J=1$           & 26 & 20 & 3 & 0.01  & 5  & 0.5 & 15 & 18\,720 \\
    \hline
    \hline
    \end{tabular}
    \caption{Numerical parameters used in the GEM calculation, with $l_{\max}=L_{\max}$ and lengths in units of $a_B$.}
    \label{tab:numparams_GEM}
\end{table}

\subsection{Results}

First, we will define the units of energy in which we will present the numerical results.
In atomic physics, energies are measured in atomic units, a.u. or Hartree units, which are equal 
to $2\text{Ry}_{\infty}$, with the Rydberg energy 
$\text{Ry}_{\infty}=m_{e} e^4 / 2 \varepsilon^2 \hbar^2$, given that 
the proton in the hydrogen model has infinite mass.
For consistency with Refs.~\cite{Combescot2019,Marsusi2022}, we introduce excitonic units, e.u. 
or excitonic Hartree units.
They are defined by twice the reduced (or exciton) Rydberg energy, 
$\text{Ry}_{M}$, taking into account the finite mass $M$ of the heavy particle (hole) as 
$\text{Ry}_{M}=\mu e^4 / 2 \varepsilon^2 \hbar^2$, where $\mu=m_{e} M /(m_{e} + M)$.
This gives us the following relation to recalculate the Rydberg (Hartree or a.u.) into the 
reduced (or exciton) Rydberg (excitonic Hartree or e.u.)
\begin{equation}
\nonumber
    \text{Ry}_{M} = \left(\frac{M}{m_e + M} \right) \text{Ry}_{\infty}.
\end{equation}
For atomic systems, the difference between the Rydberg and the reduced Rydberg energy is 
relatively small, though taking into account the finite mass of the proton shifts the hydrogen 
ground state energy from $-0.5$ a.u. to $\approx -0.4997278$ a.u.\,. In e.u. or the excitonic 
Hartree units, this energy will be independent of the masses, exactly $-0.5$ e.u.\,.
To recalculate the energy in Rydberg (Hartree or a.u.) into the energy in the reduced 
(or exciton) Rydberg (excitonic Hartree or e.u.) one can use the formula
\begin{equation}
\nonumber
    \frac{E}{\text{Ry}_{\infty}} =  \left(\frac{M}{m_e + M} \right) \frac{E}{\text{Ry}_{M}}.
\end{equation}

\begin{widetext}

\begin{table}[!h]
\begin{center}
\hspace*{-1.7cm}\begin{tabular}{cllllll}
\hline
\hline
\multicolumn{7}{c}{$J=0$} \\
\hline
 & & Our FEM solver & Our GEM solver & ThreeBSCbSSR~\cite{Liverts2013} & SWEXPH~\cite{Korobov2026}  & Combescot \cite{Combescot2019} \\
 \multicolumn{2}{c|}{$r={m_{h}}/({m_{e}+m_{h}})$} & (e.u.) & (e.u.) &  (e.u.) & (e.u.) & (e.u.) \\
\hline
\hline
 & 0 & -0.5277509 & & & -0.5277510  &  -0.5282 \\
& 0.0001 & -0.52774760 & -0.5277477274 & -0.527747729157 & -0.527747729164  &   \\
\hline
H$^{-}$ & 0.0005443206 & -0.52773300 & -0.5277331360 & -0.527733137107 & -0.527733137114  &   \\
\hline
 & 0.001 & -0.52771802 & -0.5277181948 & -0.527718196518 & -0.527718196525  &   \\
 & 0.01 & -0.5274252 & -0.5274281626 & -0.527428163600 & -0.527428163607  &   \\
 & 0.05 & -0.526234 & -0.5262523345 & -0.52625233932 & -0.52625233948  & -0.5267  \\
 & 0.10 & -0.52496 & -0.525029358 & -0.525029370 & -0.525030  & -0.5256 \\
 & 0.15 & -0.524014 & -0.524063618 & -0.524063638 & -0.524078  & -0.5247 \\
 & 0.20 & -0.52326 & -0.523341609 & -0.523341636 & -0.52341  & -0.5240 \\
 & 0.25 & -0.522846 & -0.522854649 & -0.522854683 & -0.5231  & -0.5235 \\
 & 0.30 & -0.522565 & -0.5225989 & -0.5225990 & -0.52289  & -0.5233 \\
 & 0.35 & -0.5225748 & -0.5225759 & -0.5225760 & -0.5231  & -0.5233 \\
 & 0.40 & -0.52281 & -0.522792835 & -0.522792880 & -0.5241  & -0.5235 \\
 & 0.45 & -0.523227 & -0.523263504 & -0.523263546 & -0.5253  & -0.5239 \\
 & 0.50 & -0.52399 & -0.524010130 & -0.524010140 & -0.5256  & -0.5246 \\
 & 0.55 & -0.525059 & -0.525065354 & -0.525065362 & -0.5241  & -0.5256 \\
 & 0.60 & -0.5264787 & -0.526476186 & -0.526476193 & -0.52618  & -0.5270 \\
 & 0.65 & -0.5283181 & -0.5283099 & -0.5283100 & -0.52824  & -0.5287 \\
 & 0.70 & -0.530675 & -0.530664943 & -0.530664955 & -0.530615  & -0.5310 \\
 & 0.75 & -0.533699 & -0.533689379 & -0.533689385 & -0.533623  & -0.5340 \\
 & 0.80 & -0.5376184 & -0.5376202929 & -0.5376202985 & -0.5376115  & -0.5378 \\
 & 0.85 & -0.5428769 & -0.542871317 & -0.542871323 & -0.54259  & -0.5430 \\
 & 0.90 & -0.5502789 & -0.5502762613 & -0.5502762668 & -0.55046  & -0.5503 \\
 & 0.95 & -0.5620578 & -0.562056105 & -0.562056110 & -0.56226  & -0.5621 \\
 & 0.96 & -0.5654825 & -0.56548099 & -0.56548100 & -0.56566  &  \\
 & 0.97 & -0.5695923 & -0.5695918508 & -0.5695918556 & -0.56974  & -0.5696 \\
 & 0.98 & -0.574789406 & -0.574789496 & -0.57478950 & -0.57488  & -0.5748 \\
 & 0.99 & -0.5821145 & -0.5821151234 & -0.5821151274 & -0.582153  & -0.5821 \\
 & 0.999 & -0.5956839 & -0.595685027 & -0.59568503133 & -0.59568503143  &  \\
\hline
  H$_{2}^{+}$ & 0.9994556794 & -0.597464260 & -0.597464273720 & -0.59746427506 & -0.59746427521\ldots$^{a}$  &   \\
\hline
 & 0.9999 & -0.600369 & -0.6003883 & -0.6003851 & -0.5959  &  \\
 & 1 & -0.602628 &   & & -0.5971 & -0.602633765\ldots$^{b}$ \\
\hline
\hline
\end{tabular} \\
$^{a}$ Refs.~\cite{Korobov2000,Li2007,Korobov2022} \\
$^{b}$ Ref.~\cite{Frolov2018}
\caption{\label{table1} Comparison of the $X^{+}$ trion ground-state energies (total 
angular momentum $J=0$) for different effective mass ratios $r=m_{h}/(m_{e}+m_{h})$. The energies 
are given in excitonic Hartree units (e.u.). The exciton energy in these units is exactly $-0.5$ 
e.u. for all mass ratios.
There are data obtained by our FEM and GEM solvers (second and third columns), the results of using the 
ThreeBSCbSSR program~\cite{Liverts2013} (fourth column) and the SWEXPH program~\cite{Korobov2026} (fifth column) as well as outputs of the momentum space 
calculations from Ref.~\cite{Combescot2019} (last column). Each value is given with either four decimal digits or a maximal number of common 
digits with another cell in a row plus two digits. The precision of our FEM calculation for arbitrary 
$r$ is about $10^{-4}$ e.u., whereas for $r$ corresponding to H$^{-}$ and H$_{2}^{+}$, it is below $10^{-5}$ e.u.
The precision of our GEM calculations is of about $10^{-7}$ e.u., whereas for $r$ corresponding to H$^{-}$ and H$_{2}^{+}$, it is of order $10^{-9}$ e.u.
The results for $r=0$ and $r=1$ are calculated for $r$ values shifted by $10^{-12}$.}
\end{center}
\end{table}

\subsubsection{Zero total angular momentum $J=0$}

Using FEM and GEM solvers described in Sec.~\ref{sec:methodFEM} and~\ref{sec:methodGEM}, 
respectively, we computed the eigenenergies of the trion for different effective masses of the 
constituent charge carriers. As an additional benchmark, we compared our results with the 
outcomes of three alternative approaches. A comparison of the ground-state energies of the 
$X^{+}$-trion obtained by all five methods for various mass ratios $r={m_{h}}/({m_{e}+m_{h}})$ 
is given in Table~\ref{table1}. The data are obtained for a total angular momentum $J=0$ and are 
given in excitonic Hartree units (e.u.) without rounding off.
The second and third columns give results of our FEM and GEM solvers, respectively.
The fourth column shows results of calculations using the ThreeBSCbSSR program~\cite{Liverts2013} 
which implements the numerical method based upon an expansion of the wave function over the 
Laguerre polynomials in perimetric coordinates.
The fifth column contains results using the SWEXPH program~\cite{Korobov2026} which, in turn, 
realizes a high-precision variational approach with a predefined form of the ground-state wave 
function. The last column shows results of the momentum-space calculations from 
Ref.~\cite{Combescot2019}. For a particular method, the number of converged decimal digits differs 
for different mass ratios. Consequently, there are different numbers of decimal digits in table 
cells of columns 2-5. To simplify comparison, in a particular cell the energy value is given 
with either four decimal digits or the maximal number of common digits with another cell in a 
row increased by two.

The FEM solver allowed us to achieve an accuracy comparable to that of recent calculations, in 
particular, the ground-state energy data by Combescot~\cite{Combescot2019}.
We found that the convergence of this method directly relates to the number of basis functions, 
the Legendre polynomials, over the angular variable $\theta$ in Eq.~\eqref{eq5}.
The number of finite elements (60 over each spatial variable) as well as the number of basis 
functions over the radial coordinates (damped Legendre polynomials up to the seventh order) were 
fixed high enough and, thus, less affected the precision.
The presented in column 2 results were obtained from the extrapolation of the values calculated 
for a fixed number of Legendre polynomials over the angular variable, to an infinite number of 
polynomials.
For each mass ratio, we calculated and fitted the series with up to $l=30$ Legendre polynomials 
over $\theta$. This gave us an accuracy of less than $1.3 \times 10^{-4}$~e.u. for the worst fit 
for $r \sim 0.4$ across all studied range of mass ratios.
By approaching the limiting cases $r \to 0$ and $r \to 1$, the accuracy gradually improves by, 
at least, one order of magnitude. Additionally, for the case of the positively charged molecular 
hydrogen ion H$_{2}^{+}$ we used up to $l=40$ polynomials to improve a precision of the lowest 
trion eigenenergy by, at least, two orders of magnitude, see the corresponding line in 
Table~\ref{table1}.

The results of the GEM solver appeared to be more precise than those of the FEM solver. 
The GEM energies are variational upper bounds with an accuracy better than $10^{-7}$~e.u.\ for 
all studied mass ratios and of order $10^{-9}$~e.u.\ for H$^{-}$ and H$_{2}^{+}$, see 
Table~\ref{table1}. Therefore, tracking of the results as a function of the basis size and 
further extrapolation was not required.
The results were obtained with the basis parameters listed in Table~\ref{tab:numparams_GEM}.
For the ground state calculations, the convergence is most sensitive to the number of partial 
waves taken into account, i.e. values of $l_{\max}$ and $L_{\max}$, and to the short-range 
parameter of the grid, $x_1$.
For the given $N_{\max}$ and $n_{\max}$, using only $s$~waves already produces decent, with an 
accuracy of less than $10^{-5}$~e.u., results for the ground state and a few lowest excited 
states of H$_{2}^{+}$. With $l_{\max}=L_{\max}=2$ the precision of the lowest energy values is 
improved by almost three orders of magnitude.

The method of expansion over the Laguerre polynomials in perimetric coordinates realized in the 
ThreeBSCbSSR program~\cite{Liverts2013} with 28 basis functions was employed to produce the 
ground state energy with a very good precision of eight digits for intermediate values of $r$ and 
minimum five decimal digits for very limiting case of $r=0.9999$.
For H$_{2}^{+}$, 36 basis functions were used in order to increase the accuracy even more.
One can see from Table~\ref{table1} that our FEM results and the results of the ThreeBSCbSSR 
program coincide reasonably well, with a maximal discrepancy of less than $10^{-4}$~e.u. for 
$r \sim 0.2$. Moreover, for the cases of H$^{-}$ and H$_{2}^{+}$, the agreement is by two 
orders of magnitude better.
The difference between our GEM results and the ThreeBSCbSSR data does not exceed 
$5\times10^{-8}$~e.u.\ and decreases to a few $10^{-9}$~e.u.\ for H$^{-}$ and H$_{2}^{+}$. 
It also confirms the high accuracy of the GEM.

The results of the SWEXPH program~\cite{Korobov2026} were obtained with the variational 
parameters optimized for H$_{2}^{+}$ and H$^{-}$, for those it gives the ground state energy 
with a remarkable precision of minimum 40 decimal digits. Thus, for such $r$ they are the 
reference values. For an arbitrary value of the mass ratio, the results of the SWEXPH solver 
are less superior, but still suitable for comparison.
For $r \le 0.8$, we used the ansatz optimized for H$^{-}$, whereas for other values the ansatz 
appropriate for H$_{2}^{+}$.
The reason for choosing this value of $r$ is that, at $r\sim 0.9$, as we will demonstrate below, 
the upper bound state levels appear in the spectrum, resulting in a significant change to the 
internal structure of the wave function.
The low precision of the SWEXPH program for $r\to1$ can also be caused by this effect.
For H$^{-}$ and H$_{2}^{+}$, our FEM results and the data from the SWEXPH agree within an 
accuracy of order of $10^{-7}$ and $10^{-8}$~e.u., respectively.  For intermediate values of $r$, 
the difference is larger, reaching its maximal value of about $2\times 10^{-3}$~e.u. in the 
middle of $[0,1]$, for $r\sim 0.45$.
Our GEM results for H$^{-}$ and H$_{2}^{+}$ coincide with SWEXPH results with an 
excellent precision of $4\times10^{-9}$~e.u.\,. For intermediate values of $r$, the difference 
reaches $2\times10^{-3}$~e.u., that is comparable with the difference between FEM and SWEXPH.
Similar to a comparison to our data, the discrepancy between the SWEXPH outcomes 
and ThreeBSCbSSR data is almost negligible for H$^{-}$ and H$_{2}^{+}$, whereas it reaches the 
maximum of the same $2 \times 10^{-3}$~e.u. for $r \sim 0.45$.
As a result, we conclude that for intermediate values of the mass ratios the predefined for H$^{-}$ and H$_{2}^{+}$ variational parameters of SWEXPH are less appropriate.

The results of Ref.~\cite{Combescot2019} agree with our FEM and GEM data with an uncertainty of 
less than $10^{-4}$~e.u. for $r \ge 0.9$, but, as $r$ becomes smaller, the difference to our 
data increases reaching its peak discrepancy of about $10^{-3}$~e.u. for $r \sim 0.15 - 0.40$.
It seems that for such mass ratios the method by Combescot overestimates the ground-state 
binding energies. Thus, for such $r$ the momentum-space calculations may be less precise.

\begin{table}[!h]
\begin{center}
\begin{tabular}{lllll}
\hline
\hline
 \multicolumn{5}{c}{$J=0$} \\
 \hline
 $\nu$ (vibr. q. n.)  & Our FEM result & Our GEM result & High-accuracy results & ThreeBSCbSSR~\cite{Liverts2013} \\
   & (a.u.) & (a.u.) & (a.u.) & (a.u.)  \\
\hline
\hline
 0  & -0.597139049 & -0.5971390616 &  -0.597139063123\ldots & -0.5971390629  \\
 1  & -0.5871546 & -0.5871556771 &  -0.587155679212\ldots &  -0.587155665 \\
 2  & -0.57775129 & -0.5777519014 &  -0.57775190459547 &  -0.57775178  \\
 3  & -0.56890807 & -0.5689084930 &  -0.56890849896677 & -0.5689078  \\
 4  & -0.5606091931 & -0.5606091963  & -0.56060922113307 & -0.5606062 \\
 5  & -0.5528421 & -0.552840656 &  -0.55284075021966 & -0.552840676  \\
 6  & -0.5455956 & -0.54559236 &   -0.54559265134900 & -0.545508  \\
 7  &  -0.538862 & -0.5388557 & -0.53885738734702 & -  \\
 8  &  -0.5326377 & -0.5326253 & -0.53263037975264 & -  \\
 9  &  -0.526922 & -0.52689 & -0.52691012442161 & -  \\
 10  &  -0.52172 & -0.521661 & -0.52169836942035 & -  \\
 11  &  -0.517027 & -0.5169 & -0.51700036567716 & -  \\
 12  &  -0.512867 & -0.51267 & -0.51282520352719 & -  \\
 13  &  -0.50925 & -0.5089 & -0.50918624872335 & -  \\
 14  &  -0.50621 & -0.5057 & -0.50610168128651 & -  \\
 15  &  -0.50378 & -0.50323 & -0.50359508526777 & -  \\
 16  &  -0.50198 & -0.50137 & -0.50169577359316 & -  \\
 17  &  -0.50082 & -0.50022  & -0.50043704058981 & - \\
 18  &  -0.5002 & -0.49977  & -0.49983743207558 & - \\
 19  &  -0.5001 & -0.499726  & -0.4997312306558 & - \\
\hline
\hline
\end{tabular}
\caption{\label{table2} Calculated energies of the states of H$_{2}^{+}$ with the vibrational 
quantum numbers $\nu=0-19$  for the total angular momentum $J=0$ in comparison with other 
results. The energies are given in a.u.\,. The high-precision results for 
$\nu =0$ are from Refs.~\cite{Korobov2000,Li2007,Korobov2022}, for $\nu =1$ from 
Ref.~\cite{Li2007}, and for higher energies they are from Ref.~\cite{Hilico2000}. The energies in the last column were obtained by the ThreeBSCbSSR program using 36 basis functions. For $\nu \ge 7$, results of the program did not converge.}
\label{t2}
\end{center}
\end{table}

Table~\ref{table2} shows a comparison of our results on the vibrational energy levels of 
H$_{2}^{+}$ with the high-precision data from different publications and the outputs from the 
ThreeBSCbSSR program. The energies for a full set of quantum numbers $\nu=0-19$ are given in 
a.u.\,. The distribution of the finite elements over radial variables in the FEM solver was 
optimized for the ground state calculations.
Therefore, for the lower energy levels, the FEM shows a good agreement with alternative 
approaches, with an accuracy of less than $10^{-6}$ a.u.\,. As $\nu$ increases, the discrepancy 
with the reference values gradually grows.
It reaches its maximum of $4 \times 10^{-4}$ a.u. for the last bound state with $\nu=19$.
Our GEM results are more precise than the FEM data for $\nu\le8$ and $\nu\ge17$.
For $9\le\nu\le16$ the FEM values are closer to the reference values, but lie below the exact 
energies due to the extrapolation. It is worth noting that the FEM energies for fixed numbers of 
the Legendre polynomials over the angular variable $\theta$ are above the reference values.
The GEM basis was also optimized for the ground state, which deviates from the reference 
by $1.6\times10^{-9}$~a.u. The difference grows with $\nu$, up to $3.6\times10^{-4}$~a.u.\ for 
$\nu=15$. With further increase of $\nu$, the deviation becomes gradually smaller, though, 
relative to the binding energy, the uncertainty of GEM data grows monotonically.
For the last calculated state, $\nu=19$, the accuracy is about $5 \times 10^{-6}$~a.u.\,. One 
could see that the obtained energy lies slightly above the three-body threshold. These deviations 
stem from the density of the basis covering intermediate distances: increasing the minimum range 
($x_1=0.02$) yields all 20 levels bound, at the cost of a less accurate ground state (with the 
difference from the reference value of order $10^{-7}$~a.u.).
The accuracy of the calculated weakly bound states can be improved also by increasing the 
long-range parameter $y_{N_{\max}}$ and a number of Gaussians.

The ThreeBSCbSSR program with 28 basis functions allows one to obtain only the six lowest energy 
levels. The precision of the data gradually decreases with an increase in the level number. 
The increase of the size of the basis set up to 36 allows one to calculate one more level. For 
larger numbers of basis functions, the calculation time dramatically increases. The comparison of 
our results and the data from the ThreeBSCbSSR program with 36 basis functions shows a good 
agreement of lower energy levels. The advantage of our methods is that they allow accurate 
calculations of all excited bound states.

Figure~\ref{fig:EnergyDependenceOnMassRatio} shows the calculated energy levels of the $X^{+}$
trion for the total angular momentum $J=0$ as a function of the mass ratio $m_h/m_e$.
It spans several orders of magnitude: from very large values, including the mass ratio of 
$1836.152701$ for H$_{2}^{+}$, down to almost those values for which the system is similar to 
H$^{-}$. In fact, the data for H$^{-}$ is not shown because there is only one bound state with 
the energy given in Table~\ref{table1}. As in the table, the energies in the figure are given in 
excitonic Hartree units (e.u.).
The data from the FEM and GEM calculations are precise enough to visually overlap in the plot.
The dashed horizontal line shows the three-body scattering threshold, defined exactly by the 
exciton ground-state energy of $-0.5$ e.u.\,.

\begin{figure}[h!]
\includegraphics[width=0.95\columnwidth]{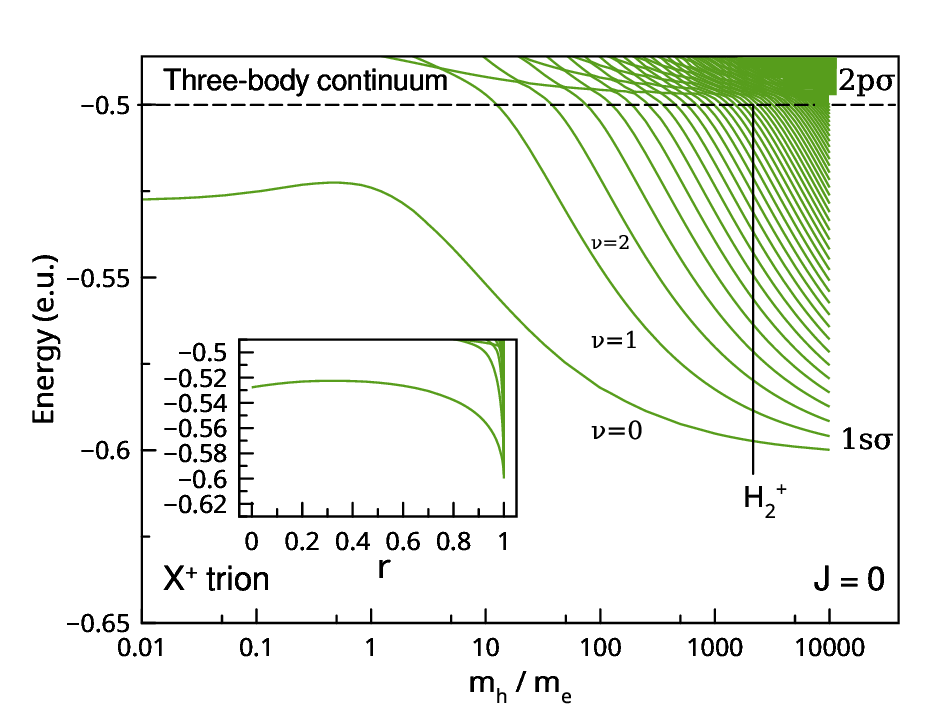}
 \caption{Calculated energy levels of the $X^{+}$ trion in a bulk crystal as a function of the effective mass ratio for a total angular momentum $J=0$. The energies are given in e.u.\,. The mass ratio corresponding to the case of the molecular hydrogen ion H$_{2}^{+}$ is denoted by a vertical line. The dashed horizontal line is the lower boundary of the continuum spectrum. The energy levels originating from the lowest ($1s\sigma$ and $2p\sigma$) subbands are shown. The quantum numbers $\nu$ of three lowest vibrational levels are given. The inset shows the energies as a function of the ratio $r=m_h/(m_e+m_h)$.}
\label{fig:EnergyDependenceOnMassRatio}
\end{figure}

For a mass ratio $m_h/m_e \lesssim 10$, there is only one bound state. This is related to the 
significant repulsion between the particles of the same charge, whose contribution, with respect 
to the kinetic energy, increases as $m_h/m_e \to 0$, see the discussion in Sec.~\ref{Sect2C}.
An appearance of the only bound state for a H$^{-}$-like system with $m_{e}/m_{p} \lesssim 0.21$
was rigorously shown by Hill~\cite{Hill1977}. However, the proved estimate of the mass ratio is 
not an exact bound.
Our calculations show that the first excited bound state disappears already for $m_h/m_e 
\approx 9.7$, i.e., when the two identical holes are about one order of magnitude heavier than 
the electron. In the model of $X^{+}$ trion, its two holes play the role of the electrons in 
H$^{-}$, whereas its electron plays the role of the proton.

For larger $m_h/m_e$, more energy levels of bound states appear in the spectrum. In contrast to 
the one-dimensional case (see discussion in Sec.~\ref{Sect2C}), there are no pairs of energy 
levels here. This originates from the fact that in 3D the coordinates of the relative motion are 
only nonnegative.
Thus, the Born-Oppenheimer potential curves have only a single QWl and the levels are 
single.

As the mass ratio increases, more vibrational energy levels, originating from the hole-hole 
vibrations, appear in the spectrum. They correspond to the lower electronic potential-energy 
curves. The energy levels originating from the lowest, $1s\sigma$, and upper $2p\sigma$ subbands 
are clearly seen in the figure. The vibrational energy levels are indexed by $\nu=0,1,2\ldots$.
The $1s\sigma$ states are characterized by an even ('gerade') wave function, and they are the 
bound states. The $2p\sigma$ subband is within the three-body continuum and thus beyond the scope 
of the current study. Whether these levels are proper resonances or numerical artifacts due to 
rigid-wall boundary conditions requires a dedicated analysis based on the complex-scaling 
method~\cite{Moi98,Belov2022}.

As it was shown by Bethe and Salpeter~\cite{Bethe1957}, for $m_h/m_e \to \infty$ the vibrational 
energies scale as $m_h^{-1/2}$. Assuming this dependence as a leading term of the asymptotic 
expansion over a small parameter $\sqrt{1-r}$, we can extend a series introduced in 
Ref.~\cite{Combescot2019} as
\begin{equation}
\label{fit1}
    E(r)=E(1)+A\sqrt{1-r}+B(1-r)+C (1-r)^{3/2}+D(1-r)^{2}+\ldots \;.
\end{equation}
This series represents a deviation of the energy from a constant value $E(1)$ as $r$ becomes 
smaller. From Fig.~\ref{fig:EnergyDependenceOnMassRatio}, one may surmise that the number of 
reasonable values of the expansion is defined by the form of a particular vibrational curve.
The analysis shows that the first five terms of this series approximate the ground state energies 
with an average accuracy of about $10^{-4}$~e.u.\,.
Interestingly enough, the upper vibrational curves are approximated by one order of magnitude better because they are localized in the vicinity of $r=1$.
The results of the fits to the most precisely calculated data are shown in Table~\ref{table3}.
There are the first five parameters of Eq.~\eqref{fit1}, including $E(1)$, fitted for the six lowest energy curves.
The fitted values are statistically significant. 
The allowed ranges of $r$ are also given.

One can see in Table~\ref{table3} that, up to a precision of calculation, the energy $E(1)$ is 
constant for all the fitted vibrational levels. This means that all the levels converge to the 
same value as $r \to 1$ (or $m_h / m_e \to \infty$), i.e. to the ground state of H$_{2}^{+}$ with 
infinite nucleus mass.
Moreover, our extrapolated energies are close to the calculated value in Table~\ref{table1}, as 
well as to the precise value of -0.602633765597 a.u., reported in Ref.~\cite{Frolov2018}. This 
additionally ensures correctness of the applied expansion.

\begin{table}[!h]
\begin{center}
\begin{tabular}{ccccccc}
\hline
\hline
 \multicolumn{7}{c}{$J=0$} \\
\hline
 $\nu$ & $E(1)$ & $A$ & $B$ & $C$ & $D$ & allowed $r$ \\
\hline
\hline
 0 & $-0.602726$ & $0.230422$ & $-0.250908$ & $0.168590$ & $-0.073041$ & $r \in [0,1]$ \\
 1 & $-0.602629$ & $0.680459$ & $-1.353728$ & $1.354479$ & $-2.013583$ & $r>0.907$ \\
 2 & $-0.602611$ & $1.132059$ & $-3.544812$ & $4.951784$ & $-14.60841$ & $r>0.97$ \\
 3 & $-0.602638$ & $1.589184$ & $-7.049382$ & $15.86373$ & $-70.56026$ & $r>0.985$ \\
 4 & $-0.602640$ & $2.043442$ & $-11.59381$ & $32.83226$ & $-187.5359$ & $r>0.99$ \\
 5 & $-0.602654$ & $2.500244$ & $-17.45588$ & $64.35537$ & $-464.5818$ & $r>0.994$ \\
\hline
\hline
\end{tabular}
\caption{\label{table3} Results of the fit of Eq.~\eqref{fit1} to the calculated data in Fig.~\ref{fig:EnergyDependenceOnMassRatio}. The first five terms of the series are used. The last column shows the allowed values of the ratio $r=m_h/(m_e+m_h)$, for which the approximation is reasonable. The values are given in e.u.\,. The fitted values of energy in e.u. for $r=1$ coincide with the values in a.u.\,. All obtained values are statistically significant.}
\label{t3}
\end{center}
\end{table}


\subsubsection{Total angular momentum $J=1$}

\begin{figure}[h!]
\includegraphics[width=0.95\columnwidth]{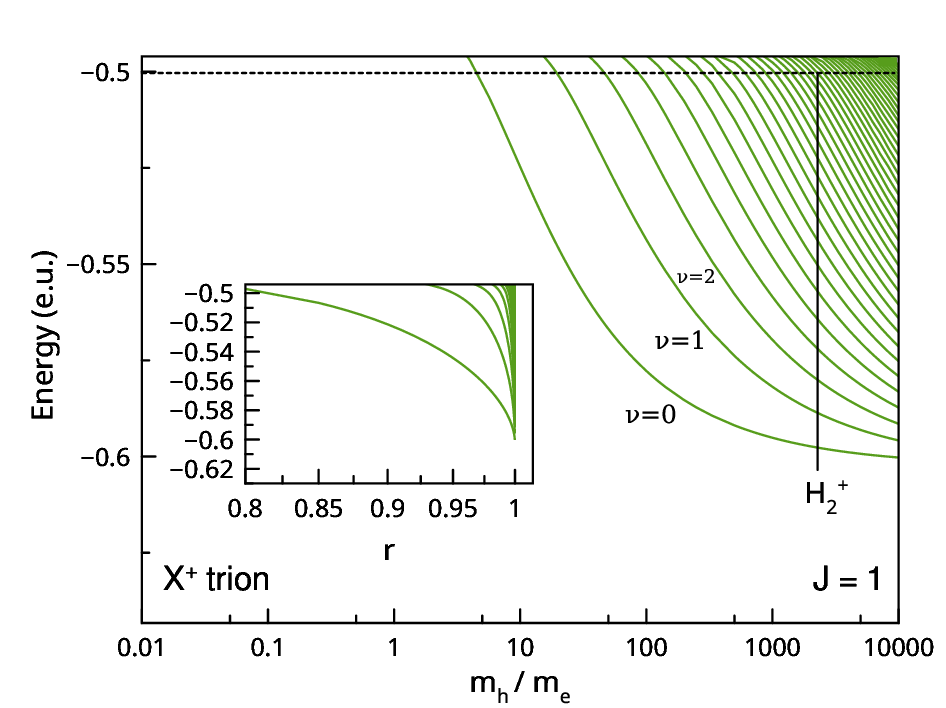}
 \caption{Calculated energy levels of the $X^{+}$ trion in a bulk crystal as a function of the effective mass ratio for the total angular momentum $J=1$. The energies are given in e.u.\,. The mass ratio corresponding to the case of the molecular hydrogen ion H$_{2}^{+}$ is denoted. The dashed horizontal line is the lower boundary of the continuous spectrum. The inset shows the energies as a function of the ratio $r=m_h/(m_e+m_h)$.}
\label{fig:Jeq1EnergyDependenceOnMassRatio}
\end{figure}

We calculated the energy levels of the trion
states for the total angular momentum $J=1$ as 
well. Due to the increased number of coupled equations and partial waves taken into 
account, calculations for $J=1$ required significantly more computational resources to store 
the matrices and to solve the eigenvalue problem.

For the FEM solver, because of limitations introduced by PARDISO to a number of nonzero 
elements of the matrices, the number of finite elements over each spatial variable was taken to 
be 30. It was only possible to achieve a convergence of the results with an accuracy of 
$4 \times 10^{-5}$ e.u. for mass ratios $r \gtrsim 0.9$. For lower mass ratios, the precision 
of our calculation is by one order of magnitude worse.
In particular, the convergence of the lowest energy for $0.1 \lesssim m_{h}/m_{e} \lesssim 5$, 
when, as it will be clear below, the ground state wave function is large enough, requires an 
increased number (up to 40) of Legendre polynomials over the angular variable.
To compensate this increase and keeping the same matrix sizes, the number of finite elements over 
the spatial variables was decreased down to 20. As a result, for 
$0.1 \lesssim m_{h}/m_{e} \lesssim 5$, the precision of our FEM calculation is $3 \times 10^{-3}$ 
e.u. For lower mass ratios, including the one corresponding to H$^{-}$, our precision is again 
below $10^{-3}$ e.u.\,.

For the GEM calculations, memory constraints also restricted the sizes of the matrices. 
The larger number of $(l,L)$ channels, $n_{lL}=9$, required for $J=1$, forced us to use fewer 
Gaussians per radial coordinates than for $J=0$.
Nevertheless, the precision of GEM results for H$_{2}^{+}$ ground-state energy is of order 
$10^{-8}$~e.u.\,.
For $0.99 \le r \le 0.999$ the estimated accuracy is by one order of magnitude worse
and for $0.85< r \le 0.98$ it is of order $10^{-6}$~e.u.
For smaller values of $r$, the calculated ground state is either very weakly bound or completely 
delocalized: its wave function extends beyond the basis range. 
Thus, the precision of $2\times10^{-5}$~e.u. is estimated for $r \sim 0.81$.

The calculated energy levels for the total angular momentum $J=1$ as a function of $m_{h}/m_{e}$ 
are shown in Fig.~\ref{fig:Jeq1EnergyDependenceOnMassRatio}. Similar to the case of $J=0$, for 
large mass ratios there are many vibrational energy levels. As $m_{h}/m_{e}$ decreases, the 
energies of the bound states gradually increase. Therefore, the bound state energies transit into 
the continuous part of the spectrum. The number of bound states decreases and for 
$m_{h}/m_{e} < 4.2$
there are no bound states in the system at all.
The achieved precision of minimum $3 \times 10^{-3}$ e.u. for FEM calculations and, 
moreover, of order $10^{-5}$ e.u. for GEM calculations allows us to reasonably argue 
about the absence of bound states for such mass ratios. This additionally confirms the well-known 
fact that H$^{-}$-like systems have only one bound state~\cite{Hill1977}, corresponding to the 
case $J=0$.
In fact, this means that for semiconductor compounds with effective masses of the same order of 
magnitude, for instance, for Cu$_{2}$O, there is only one trion state with $J=0$.
All the energies calculated by our methods for $m_{h}/m_{e} < 4.2$ are in the three-body 
continuum. As it is unclear whether they correspond to proper resonant states or artificially 
discretized continuum states, they are beyond the energy range shown in 
Fig.~\ref{fig:Jeq1EnergyDependenceOnMassRatio}. In any case, they lie far enough above the 
scattering threshold. This can be seen by a narrow strip of the continuum region above the dashed 
line in Fig.~\ref{fig:Jeq1EnergyDependenceOnMassRatio}.

The comparison of the energies of the lowest state of the $X^{+}$ trion with total angular 
momentum $J=1$ is given in Table~\ref{tableJ1}.
The data are only given for values $r \ge 0.8$,
since for most of these values both methods find a bound state.
The second and third columns show our results, whereas the fourth column lists the data from the 
SWEXPH program~\cite{Korobov2026}.
The energy values are given with the number of decimal digits corresponding to the order 
of the estimated uncertainty plus two digits, but not less than four digits.
For the mass ratio corresponding to H$_{2}^{+}$, the additional high-precision result from 
Ref.~\cite{Hilico2000} is given in e.u. as a reference.
The results of SWEXPH are obtained with the variational parameters optimized for the ground state 
of H$_{2}^{+}$ with $J=0$.

It can be seen that, similar to zero total angular momentum data, the GEM results are more 
accurate than the FEM results. For the mass ratio corresponding to H$_{2}^{+}$, the results of 
GEM and the SWEXPH program remarkably agree up to the eighth decimal digit.
Moreover, they coincide with the reference value from Ref.~\cite{Hilico2000} with a similar 
precision of $10^{-8}$ e.u.\,.
The FEM results are by three orders of magnitude less precise.
For smaller values of $r$, the discrepancy between FEM and GEM data gradually increases as 
the mass ratio approaches $r \sim 0.81$. For this $r$, both the results agree with an uncertainty 
of about $3 \times 10^{-4}$ e.u.\,.
For such ratios, the wave function extends very far from the origin and further, as the energy 
crosses the scattering threshold, becomes completely delocalized. A calculation of weakly 
bound states requires using the extended angular basis and an increased value of the parameter 
$y_{N_{\max}}$ to achieve convergence in our FEM and GEM computations, respectively.
In the other limit, $r \to 1$, the FEM and GEM results agree well for $r=0.9999$, but the 
SWEXPH program gives much less precise energy.
In Table~\ref{tableJ1} the energy for the mass ratio $r=1$ is obtained for $r=1-10^{-12}$.
The value is slightly above the related energy for $J=0$.

It is worth noting that for small mass ratios, $m_{h}/m_{e} \lesssim 1$ or $r \lesssim 0.5$, the 
SWEXPH program gives the energy corresponding to the discrete part of the spectrum. In 
particular, for the negatively charged hydrogen ion H$^{-}$, it yields the energy of about 
$-0.5071$~e.u.\,. Both our FEM and GEM results do not show any bound states for 
$m_{h}/m_{e} < 4.2$. All our results are in the continuous part of the spectrum. Therefore, we 
believe that the SWEXPH program with the predefined form of the ground-state wave function provides only approximate data for this range of the mass ratios.

\begin{table}[!h]
\begin{center}
\hspace*{-1.7cm}\begin{tabular}{clllll}
\hline
\hline
\multicolumn{6}{c}{$J=1$} \\
\hline
 & & Our FEM result & Our GEM result & SWEXPH~\cite{Korobov2026} & Reference value  \cite{Hilico2000}  \\
 & $r={m_{h}}/({m_{e}+m_{h}})$ & (e.u.) & (e.u.) & (e.u.) & (e.u.) \\
\hline
\hline
 & 0.80 & -0.49927 & -0.4991609  & -0.4883  & \\
 & 0.81 & -0.50054 & -0.5002221  & -0.4980  & \\
 & 0.85 & -0.50762 & -0.50729494 & -0.5056  & \\
 & 0.90 & -0.521762 & -0.52164086 & -0.5235  & \\
 & 0.95 & -0.544489 & -0.54447612  & -0.5462  & \\
 & 0.96 & -0.550755 & -0.55074404  & -0.5522  & \\
 & 0.97 & -0.557957 & -0.55796718  & -0.5590  & \\
 & 0.98 & -0.566572 & -0.56659133  & -0.5672  & \\
 & 0.99 & -0.577706 & -0.577730515  & -0.57789  & \\
 & 0.999 & -0.595174 & -0.595202342 & -0.595202318  & \\
\hline
  H$_{2}^{+}$ & 0.9994556794 & -0.597170 & -0.5971987935 & -0.5971987953 & -0.5971988064 \\
 & 0.9999 & -0.600308 & -0.600338716  & -0.5959 & \\
 & 1 & -0.602479 &  &  \\
\hline
\hline
\end{tabular}
\caption{\label{tableJ1}Comparison of the $X^{+}$ trion lowest-state energies (the total angular momentum $J=1$) for different effective mass ratios $r=m_{h}/(m_{e}+m_{h})$. The energies are given in excitonic Hartree units (e.u.). For the mass ratio corresponding to H$_{2}^{+}$, the high-precision data from Ref.~\cite{Hilico2000} in a.u. is additionally given. The energy for the mass ratio $r=1$ is obtained from the extrapolation. For $r=0.81$ the GEM calculation used an extended basis with $y_{N_{\max}}=129$. The energy for $r=1$ is calculated for $r$ value shifted by $10^{-12}$.}
\end{center}
\end{table}
\end{widetext}

\section{Conclusion and outlook}
In this article, we calculated several of the lowest energy levels of the semiconductor trion 
for a wide range of hole-to-electron mass ratios, covering most of the existing semiconductor 
compounds. We considered trions in a quantum wire as well as in a bulk crystal.
We obtained qualitatively unified dependencies of the bound-state energies on $m_h / m_e$ and 
observed changes of the energy spectra during a crossover from small to large mass ratios.

The trion energy levels in the bulk were computed by two different methods, FEM and GEM, 
which produced similar results. We show the advantage of the GEM: its results are more precise 
than the FEM data.
For both the 1D and 3D geometries of the system there is only one bound state for the 
small values of the mass ratio, as predicted for the analogous atomic H$^{-}$ 
system~\cite{Hill1977}.
Our calculations improve the precision of the data from 
Refs.~\cite{Usukura1999,Combescot2019,Marsusi2022} and demonstrate that for most semiconductor 
compounds only one trion state with zero total angular momentum exists. The sole trion state 
takes place for the mass ratio of two different particles, i.e. $m_h/m_e$ in case of $X^{+}$ 
trion and $m_{e}/m_{h}$ for $X^{-}$, to be below $9.7$.
For $J=1$, both applied methods show no trion bound states for the ratio below $4.2$.
With increasing mass ratio, more three-body bound states originating from the hole-hole 
vibrations appear in the spectrum.
Their energies are simultaneously decreasing, gradually filling all the energy range between the 
ground state and the continuum. For $m_h / m_e \sim 1836.152701$, the studied system is identical 
to H$_{2}^{+}$ with 20 well-known vibrational states. In the limit of $m_h / m_e \to \infty$ the 
quantization disappears and the density of three-body states forms a continuum above the lowest 
energy of the H$_{2}^{+}$ with the infinite nucleus mass.

The binding energy of the trion (depending on whether it is $X^{+}$ or $X^{-}$ trion) is of about 
$5-10 \%$ of the exciton binding energy. From our calculations one can obtain more precise values 
by scaling the numerical results for particular semiconductor compounds by their corresponding 
Hartree energies.

The results obtained here within the simplest two-band model show a unified dependence of bound 
state energies on the mass ratio.
In contrast to earlier works~\cite{Lin1995}, this dependence is nonlinear:
one can observe the crossover of the energy levels 
based on the same theoretical footing.
Nevertheless, the energies above the three-body scattering threshold, 
the three-body resonant states~\cite{Happ2025}, can be further studied using a more advanced 
complex-scaling method~\cite{Moi98,Belov2022}. Our results can also be improved for the 
particular semiconductors by taking into account the complex valence-band 
structure~\cite{Scheuler2026}, nonparabolicity of the conduction band~\cite{Belov2017} and other 
effects, i.e., the central cell corrections~\cite{Schweiner16b}, screening of the 
potential~\cite{Rytova,Keldysh}, and so on.

\acknowledgements
The authors are grateful to I.~N.~Filikhin, R.~Ya.~Kezerashvili, and V.~I.~Korobov for 
fruitful discussions. P. A. B. and S. S. acknowledge financial support by the National Research Council Canada, grant No. QSP-129-1.

\bibliography{paper}

\end{document}